\documentclass[aps,pra,twocolumn,superscriptaddress,showpacs,10pt]{revtex4-2}

\usepackage{amsmath,amssymb}
\usepackage{graphicx}
\usepackage{hyperref}
\usepackage{braket}
\usepackage{color}
\usepackage{tikz}
\usepackage{chemfig}
\usepackage{bm}
\usepackage{siunitx}
\usepackage{booktabs}
\usepackage{multirow}
\usepackage{algpseudocode}
\makeatletter
\newcounter{algorithm}
\renewcommand{\thealgorithm}{\arabic{algorithm}}
\newcommand{\ftype@algorithm}{4}
\newcommand{\fnum@algorithm}{\textbf{Algorithm~\thealgorithm}}
\newcommand{\ext@algorithm}{loa}
\newenvironment{algorithm}[1][tbp]{%
  \@float{algorithm}[#1]%
  \hrule height.4pt depth0pt%
  \setlength{\abovecaptionskip}{3pt}%
    \setlength{\belowcaptionskip}{1pt}%
  \def\@algorithm@aftercaption{\kern\belowcaptionskip\hrule\kern2pt}%
  \let\@old@caption\caption
  \renewcommand{\caption}[1]{\@old@caption{##1}\@algorithm@aftercaption}%
}{%
  \kern2pt\hrule\relax%
  \end@float
}
\makeatother

\usepackage{pgfplots}
\pgfplotsset{compat=1.18}
\usetikzlibrary{arrows.meta, decorations.pathmorphing}

\newcommand{\ketbra}[2]{\left| #1 \right\rangle \left\langle #2 \right|}

\begin{document}

\title{A unified quantum computing quantum Monte Carlo framework through structured state preparation}

\author{Giuseppe Buonaiuto}
\affiliation{Fujitsu Research of Europe Ltd., Pozuelo de Alarcón, 28224 Madrid, Spain}

\author{Antonio Márquez Romero}
\affiliation{Fujitsu Research of Europe Ltd., Pozuelo de Alarcón, 28224 Madrid, Spain}

\author{Brian Coyle}
\affiliation{Fujitsu Research of Europe Ltd., Slough SL1 2BE, UK}
\affiliation{School of Informatics, University of Edinburgh, Edinburgh, UK}

\author{Annie E. Paine}
\affiliation{Fujitsu Research of Europe Ltd., Slough SL1 2BE, UK}

\author{Vicente P. Soloviev}
\affiliation{Fujitsu Research of Europe Ltd., Pozuelo de Alarcón, 28224 Madrid, Spain}

\author{Stefano Scali}
\affiliation{Fujitsu Research of Europe Ltd., Slough SL1 2BE, UK}
\affiliation{University of Exeter, Department of Physics and Astronomy, Stocker Road, Exeter EX4 4QL, UK}

\author{Michal Krompiec}
\affiliation{Fujitsu Research of Europe Ltd., Slough SL1 2BE, UK}

\date{\today}

\begin{abstract}

We extend Quantum Computing Quantum Monte Carlo (QCQMC) beyond ground-state energy estimation by systematically constructing the quantum circuits used for state preparation. Replacing the original Variational Quantum Eigensolver (VQE) prescription with task-adapted unitaries, we show that QCQMC can address excited-state spectra via Variational Fast Forwarding and the Variational Unitary Matrix Product Operator (VUMPO), combinatorial optimization via a symmetry-preserving VQE ansatz, and finite-temperature observables via Haar-random unitaries. Benchmarks on molecular, condensed-matter, nuclear-structure, and graph-optimization problems demostrate that the QMC diffusion step consistently improves the energy accuracy of the underlying state-preparation method across all tested domains. For weakly correlated systems, VUMPO achieves near-exact energies with significantly shallower circuits by offloading optimization to a classical tensor-network pre-training step, while for strongly correlated systems, the QMC correction becomes essential. We further provide a proof-of-concept demonstration that Haar-random basis state preparation within QCQMC yields finite-temperature estimates from pure-state dynamics.

\end{abstract}

\maketitle

\section{Introduction}

Simulating and characterizing complex classical and quantum systems, ranging from large networks to exotic quantum materials~\cite{Donnelly2026}, is likely to be one of the most important goals of modern science and technology. However, the complexity itself makes the realization of functional and expressive algorithms particularly challenging, especially in the presence of strong correlations or when the problem size grows significantly~\cite{Fan2025}. When the focus of the study falls on chemical or material science systems, the family of Quantum Monte Carlo (QMC) methods represents a common yet successful path for estimating their spectral and dynamical information~\cite{Venegas,petruzielo2012semistochastic,Hu2006,Austin2012}. However, an efficient deployment of these methods faces several hurdles: their scaling properties can be limited, and more importantly, their performances are hindered by the sign problem~\cite{PAN2024879}, which becomes particularly relevant for deeply quantum problems, as in strongly correlated electronic structures and nuclear physics~\cite{carlson2015}. Full Configuration Interaction Quantum Monte Carlo (FCIQMC)~\cite{booth2009fermion,cleland2012taming}, is regarded as one of the most interesting algorithms among the QMCs as it foster an incremental enhancement of the accuracy of the estimation, by sampling relevant wavefunctions out of an entire sector of the Hilbert space of the given system. In particular, it has been demonstrated that in FCIQMC, through accurate state basis selection, the sign problem is controllable and can indeed be mitigated, compared to other methods~\cite{Kolodrubetz}. Even though FCIQMC is still a fundamental reference for the field, in its purely classical state implementation, it possesses a limitation: the so-called walkers--i.e., the set of states that participate in the ``game of life", when classically constructed, do not retain enough expressive power to deal with complex quantum scenarios. For fermionic systems in particular -- where, usually, the wavefunctions are made of Hartree-Fock states or selected configuration interaction -- this fact makes the sign problem hard to mitigate~\cite{troyer2005}. In the attempt to regularize results, the amount of resources, both in terms of time steps and number of walkers, which are necessary to reach convergence and stabilize the sign, grows fast, making the entire algorithm potentially infeasible due to excessive memory consumption. Even though many methods have been developed to deal with these problems~\cite{BoothJCP,GhanemJCP}, there is not a general recipe nor a unique consensus on how to tackle them. 

Alongside QMC methods, the development of quantum algorithms promises to partially mitigate the challenges of purely classical approaches to system simulations~\cite{Daley2022}. Although they are improving at an increasingly fast rate, we are still in the early stage of their development: thus, a set of hybrid variational quantum algorithms (VQA) is on the rise~\cite{Qi2024}, as they promise to be useful in some of the instances aforementioned~\cite{eisert2025mind}, by leveraging the quantum features embedded in some core structures of the computation. However, these algorithms face several intrinsic challenges too, which limit their practical applicability, including: lack of guarantees of convergence, the emergence of barren plateaus~\cite{Larocca2025}, wherein the learning landscapes flatten, preventing the convergence towards the target minima, and obviously the detrimental effects coming from quantum noise~\cite{Saib_proc}. Even if recent works are attempting to mitigate these problems, overcoming them completely remains complicated, especially when the goal is obtaining highly accurate estimations.

The so-called Quantum Computing Quantum Monte Carlo (QCQMC) frameworks~\cite{kanno2024quantum,qcqmc_tp,Huggins2022} are a set of promising examples of synergy between QMC and VQA that serve as a possible way forward for overcoming their individual limitations. In particular, in Ref.~\cite{qcqmc_tp}, the authors solved the expressivity challenges of FCIQMC walkers by generating quantum states via a quantum circuit obtained from the variational quantum eigensolver (VQE). As a direct effect, stemming from the methodological way of improving the initial overlap between the trial and the target state, this approach limits the explosion of the population of walkers in the QMC computation. On the other hand, by employing shallow versions of VQE, problems like barren plateaus and vanishing gradients are largely circumvented. Together with the aforementioned variational algorithm approach, other hybridization strategies have been adopted. For instance, tensor-network based approaches, already well-known in the purely classical QMC regime~\cite{VidalTNqmc,ChenJielun}, are worth mentioning. In particular, in Ref.~\cite{JiangTong}, the use of MPS-based circuits in a QMC pipeline was probed, opening the way of using these resource efficient quantum ansatz for tackling complicated electronic structure calculations. Most recently, the hybrid tensor network-QMC~\cite{kanno2024quantum} demonstrates strategies for achieving higher accuracy than classical QMC, while avoiding complex computational costs encountered in the optimization of VQE circuits in QCQMC. 

In light of these encouraging developments, some interesting questions emerge: in particular, the possible broader applicability of the QCQMC algorithm, away from the ground state energy estimation, remains unexplored, contrary to classical QMC and VQE, both of which have been applied in  various scenarios~\cite{Shepard2025,Higgott2019}. This possibility itself opens another question: VQE generated states not only can be expensive, but also inadequate for tackling the specific problem too. Hence, answering the question on the optimal way of constructing quantum states for the specific problem instance is of crucial relevance. 

In this work, we address these open questions by proposing a general workflow for applying QCQMC to different domains of interest. We achieve this versatility by methodically
varying the unitary operator, which generates a domain-specific quantum basis. Moreover, the Monte Carlo specifications and the estimation strategies for the target quantities are adapted to deal with each and every scenario considered. The basis state generation techniques in particular are here selected with the idea of \textbf{(i)  extending } the native proposal of QCQMC with VQE to other class of problems (such as in optimization), \textbf{(ii) expanding }it to larger portions of the spectrum of quantum systems, \textbf{(iii) reducing} the overall quantum deployment cost without sacrificing performance, \textbf{(iv) broadening the scope} to thermal averages, in line with the underlying formulation of FCIQMC as a first order expansion of a thermalization process~\cite{DeGrandi}.  Specifically, each of the aforementioned situations will be addressed in the following, considering rather standard examples in quantum simulations, ranging from chemistry, nuclear models, to MaxCut optimization. Note that we will not be addressing the sign problem in each scenario directly, but rather, we will observe its substantial mitigation in the convergence of the QMC trajectories. 

The remainder of this paper is organized as follows. In Section~\ref{sec:methods} we provide the general step and characteristics of the cross-domain QCMQC pipeline. Moreover, we describe the specific minimal quantum circuit modifications, with respect to the seminal work, for estimating the relevant quantities in the Monte Carlo evolution. Furthermore, in Sec.~\ref{sec:unitary}, we provide an overview of each basis preparation technique employed in the work, most of which are defined by variational quantum algorithms. In Sec.~\ref{sec:hamiltonians}, the Hamiltonian encoding for each problem tackled with the cross-domain QCQMC is described. We then present benchmark results for applying the specific variations of the cross-domain QCQMC for each specific problem of interest (Sec.~\ref{sec:results}). Finally, in Sec.~\ref{sec:conclusion}, we  present a concise discussion of the key findings, highlighting possible further lines of investigations.
\section{Cross domain QCQMC}\label{sec:methods}
In this paper, we will refer to the QCQMC version by~\cite{qcqmc_tp}: QCQMC builds directly upon the well-established Full Configuration Interaction Quantum Monte Carlo (FCIQMC)~\cite{booth2009fermion,cleland2012taming} framework, which can be understood as a stochastic unraveling of the non-unitary imaginary time evolution operator~\cite{Kolotouros} $e^{-\tau H}$. This grants the asymptotic convergence to the ground state of the Hamiltonian $H$, provided a sufficiently large overlap between the initial and the ground state itself is provided. A projective modification of FCIQMC~\cite{blunt2015excited} further allows us to estimate the low-lying excited states; similarly, its density matrix equivalent~\cite{BluntDMQMC} (called DMQMC) enables us to estimate thermal averages by exploiting the dynamics of the entire density matrix of the system. In this sense, FCIQMC has been proven to possess, already in its classical state preparation version, a remarkable versatility. We provide a basic explanation of this method in Appendix~\ref{sec:fciqmc} for completeness. 

The essence of QCQMC lies in the construction of a set of \emph{quantum walker states}, i.e., a set of basis states (prepared via quantum circuits), which in principle entail structures either inaccessible directly or hard to classically simulate. Specifically, in place of the standard walker basis states $\ket{b_i}$, which are usually expressed as single Slater determinants--computational basis states under the Jordan-Wigner transformation-- in QCQMC, the walkers are constructed via quantum circuits defined via the application of a unitary operator: 
\begin{equation}\label{eq:walker_transformation}
    \ket{\widetilde{\psi}_i} = U _g\ket{b_i}.
\end{equation}
The unitary operator, $U_g$, is constructed with the aim of maximizing the overlap with the target ground state, and for boosting the expressivity of the ansatz itself. A natural choice for realizing it, then, is through the usage of a VQE optimization routine, leveraging parametrized circuits. Once the quantum walker basis is obtained, the QCQMC pipeline proceeds in analogy with FCIQMC. The only difference that emerges here is that, as the core structures are now effectively a quantum circuit, the key quantities governing their stochastic dynamic, such as the Hamiltonian transition elements, need to be evaluated via a quantum circuit as well. In the seminal work, the authors proposed a set of quantum circuits for dealing efficiently with the estimation. Here, in addition to generalizing the pipeline, especially in the state preparation sector, we add a small modification to the original circuit proposals, without altering the idea behind them. 
In the following, we describe the protocol that deals with estimating quantities from different domains using QCQMC, providing, at the same time, more details about the core functioning of the QCQMC itself. 
\subsection{The general algorithm}
\label{sec:qcqmc_protocol}
\begin{figure}
    \includegraphics[width=\linewidth]{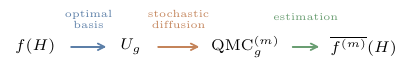}
    \caption{A schematic workflow of the Cross-domain QCQMC.}
    \label{fig:qmc_pipeline}
\end{figure}

The essential logic of the generalized protocol and pipeline for QCQMC, which we propose in this work, is depicted in Fig.~\ref{fig:qmc_pipeline}.
The aim is to estimate a certain quantity $f(H)$, related to a specific problem encoded in a Hamiltonian $H$. This quantity of interest can be the ground state energy, the entire spectrum, the optimal cut of a graph, or a finite-temperature observable. Depending on the problem, an optimal basis for finding the solution is constructed in the form of a unitary operator $U_g$, acting on a set of trial basis states. This operator can be a single transformation or a set of individual $g$ realizations $\{U_g\}$, with $g=1,\dots,G$. This set of operators can be obtained via variational optimization or can be defined \textit{ad hoc} for the specific circumstance. The new basis will form the set of quantum walkers, which then undergoes several $\text{QMC}^{(m)}_g$ Monte Carlo diffusion processes, where $m=0,\dots ,M-1$ represents the number of serialized instances of the QMC routine, i.e., the number of ordered states to be estimated, (e.g. excited states). Such processes are expected to foster the statistical estimation of the target quantities via the estimators $\bar{f}(H)$. For instance, we recover the original QCQMC formulation if the problem of interest is the ground state energy, $f(H)=E_{\rm{GS}}$ of a given system $(H)$, the optimal unitary is obtained via VQE, $U_g \equiv U=U_{\text{VQE}}$ and $\text{QMC}^{(m)}_g=\text{QMC}^{(0)}_{0}$, i.e. is a single instance of QMC for a single state. 
\begin{table*}
\label{tab:qmc-summary}
\centering
\setlength{\tabcolsep}{12pt}   
\renewcommand{\arraystretch}{1.4} 

\begin{tabular}{lccccc}
\hline
\(\mathbf{f(H)}\)
    & \(U_{g}\)
    & \(\mathbf{\,\mathrm{QMC}^{(M)}_{G}}\)
    & \(\mathbf{\overline{f^{(M)}}(H)}\)
    \\
\hline
\textbf{Ground State}               & VQE/VFF/VUMPO  & $M=0,G=1$ & $E_{\rm{pr}}^{(m)}$ &  \\
\textbf{Excited States}               & VFF/VUMPO & $M\neq0, G=1$ & $E_{\rm{var}}^{(m)}$  \\
\textbf{Optimization problem}     & VQE & $M=0, G=1$ & $E^{0}_{\rm{pr}}, |\widetilde{\Psi}^{0}\rangle$   \\
\textbf{Finite temperature averages} &  Random Haar/$t$-design & $M=0, G > 1$ & $E^{0}_{\rm{sf}}$ \\
\hline
\end{tabular}
\caption{A summary of the experiments presented in this work, listing for each problem $f(H)$, the optimal unitary quantum basis generation adopted $U_g$, the Monte Carlo  diffusion strategy QMC$^{(M)}_{G}$ and the target estimator $\overline{f^{(M)}}(H)$. Notice that for the optimization problem, the estimator for the target is given by the energy and by the reconstructed state $|\widetilde{\Psi}^{0}\rangle$.}
\end{table*}
Let us break down each part of the protocol and dive into its technical details.
\paragraph{\textbf{Hamiltonian Encoding and Symmetries.}} Once a problem, $f(H)$, is defined, the first step consists of finding a suitable Hamiltonian that describes it. For physical and chemical systems, this is trivial, as the problem itself is defined by the Hamiltonian. In optimization, for instance, things are more subtle, as the problem is formulated in another representation space and the mapping to a Hamiltonian may be non-trivial~\cite{Egginger}. Once the Hamiltonian is obtained, it may need to be transformed to a qubit representation to be deployed on a quantum circuit. Standard encoding strategies are the Jordan-Wigner~\cite{JordanWigner} and the Bravyi-Kitaev~\cite{bravyi2002fermionic} transformations or more exotic strategies such as the compact encoding~\cite{derby_compact_2021}. 
In order to compute expectation values, the Hamiltonian operator is expressed as a Linear Combination of Unitaries (LCU)~\cite{childs2012hamiltonian},
\begin{equation}\label{eq:lcu_ham}
    H = \sum_i h_i P_i,
\end{equation}
where $h_i$ are the coefficients of the combination and $P_i$ are Pauli strings, Kronecker products of the set of Pauli gates $\{I, X, Y,Z \}^{N_q}$, which can be directly implemented on a quantum circuit. While the total complexity of these Hamiltonians is of $\mathcal{O}(4^{N_q})$, where $N_q$ is the number of qubits, in many realistic cases the Hamiltonian is bounded by a polynomial complexity of $\mathcal{O}(N_q^4)$. Moreover, controlled approximations like qDRIFT~\cite{qdriftI,qdriftII,qdriftIII,qdriftIV} may also be implemented for extracting the relevant features of the Hamiltonian based on a probability distribution of its coefficients.
A further convenient step to consider before starting the QMC estimation pipeline consists in exploiting the presence of known symmetries in the Hamiltonian operator. In general, this can lead to a significant reduction in the computational cost of the properties of the systems: in fact, when a system's Hamiltonian commutes with a particular symmetry operator, the system's eigenstates can be classified according to the eigenvalues of the symmetry operator itself, inducing a block-diagonal structure of the Hamiltonian, in a suitable basis. In this fashion, the problem can be reduced into smaller symmetry-preserving subspaces. While this constitutes a general advantage, finding a set of symmetries of the system is not straightforward. When applicable, we leverage Hamming weight symmetry: a state exhibits this symmetry if the count of ``1''s in its computational basis expansion remains constant. When the system has such a property, calculation for specific state characterization in the context of QCQMC can be dramatically reduced by considering only quantum walkers with a fixed Hamming weight, i.e., quantum walkers are described in a basis with an unchanged number of ones. Clearly, the symmetry itself needs to be preserved by the action of the unitary $U_{g}$ used for generating quantum walkers. Equivalently, this fact induces a limitation in the selection of the variational ansatz for the state preparation.
\paragraph{\textbf{Quantum Walkers Preparation.}} Given the problem specifications, $f(H)$, the next step is to find a suitable set of unitary operators $U_{g}$, or quantum circuits, that generate problem-efficient quantum basis states.  For estimating ground and excited energies, we are going to look for such  basis using variational quantum algorithms whose specification will be described in the following section. The same strategy will be applied to solve optimization problems. When dealing with thermal averages, on the contrary, this state preparation step is not strictly variational, but rather deals with the construction of a unitary circuit sampled from the Haar distribution. All the variational procedures involved follow the same logic: they start with the preparation of a parametrized quantum state  $\ket{\Xi(\bm{\eta})} = U_g(\bm{\eta}) \ket{0}^{\otimes n}$, trained to optimize a certain objective function. Once the training is over, the unitary circuit $U(\bm{\eta}_{\rm{opt}})$ is used for rotating the classical walker basis, thus to construct the quantum walkers, $ |\widetilde{\psi}_i\rangle = U_g(\bm{\eta}_{\rm{opt}})|b_i\rangle$ with $i=0,\dots, \dim(H)$. In the following, we will omit the $\bm{\eta}_{\rm{opt}}$ when referring to this basis preparation step. 

\begin{figure}
    \includegraphics[width=0.8\columnwidth]{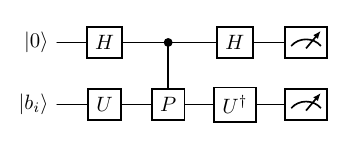}
    \caption{Modified Hadamard test to compute the absolute value of all the transition matrix elements $|\widetilde{P}_{ij}| = |\langle b_j|U^{\dagger}PU|b_i \rangle|$ for a fixed input state $|b_i\rangle$.}
    \label{fig:modHij_U}
\end{figure}

\begin{figure}
    \centering
    \includegraphics[width=0.99\columnwidth]{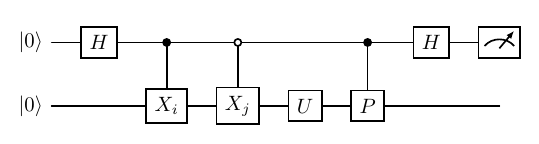}
    \caption{Modified Hadamard test to compute the transition matrix element $\widetilde{P}_{ij} = \langle b_j|U^{\dagger}PU|b_i \rangle$ and thus estimate the sign, where $X_i|0\rangle = |b_i\rangle$. }
    \label{fig:P_overlap}
\end{figure}

The rationale behind the selection of the strategy for crafting the $U_{g}$ transformation is simple and yet effective: the new basis should provide an educated guess of states with good overlap with the target states. Furthermore, the structure itself of the circuits describing such states should be rich enough to retain, to some extent, non-classical phenomena in the QMC propagation.
For instance, if the target is a large portion of the spectrum,  VQE alone fails in generating states with a robust overlap with the target excited states so approximate quantum diagonalization schemes such as \emph{variational fast forwarding} (VFF) can be the ideal variational strategy. However, VFF is an expensive routine due to the difficulty of quantumly optimizing parameters, and the increased qubit and circuit depths required. Then, one can mitigate this expense by using tensor-network pre-training techniques such as the \emph{variational unitary matrix product operator} (VUMPO). These are particularly suitable when the target has a subset of well structured and classically tractable correlations. A concrete example is where a well defined portion of the Hamiltonian generates area-law entanglement, while the general Hamiltonian admits classically intractable correlations. It is the former part that VUMPO would target, while quantum enhanced QMC would target the latter.
\paragraph{\textbf{Initialize Quantum Walker Distribution.}} The quantum walker basis for each $g$ and for a given $m$, i.e. $\lbrace U_{g}\ket{b_i}\rbrace_{i=0}^{n}=\lbrace \ket{\widetilde{\psi}^{(m)}_i}\rbrace_{i=0}^{n}$, is then fed into the dynamical pipeline. Notice that in this work multiple $U_g$ are employed only when dealing with thermal averages, i.e. in all other instances a single $U$ will define the basis preparation step. The starting point is then to make a certain number of copies $N_0$ of the initially chosen state, which we call for the sake of clarity $\ket{\widetilde{\psi}_{0}}$. In case of multiple serialized estimations,  $m>0$, i.e., for excited states, we obviously have an additional step in the pipeline, as clarified in the pseudo-code: the orthogonalization step, inherited from the classical FCIQMC. Hence, the dynamical routing of the algorithm will be initialized by a set of $\{N^{(m)}_{0}\}_{m=0}^{M}$ initial walkers, $\ket{\widetilde{\psi}^{(m)}_{0}}$, one for each target state. One of the main advantages of the pipeline proposed in this work is that we do not require $M$ different unitary rotation for tackling each target state: VFF and VUMPO allow us to prepare optimal or near-optimal starting quantum walkers with a single variational training run. 
\paragraph{\textbf{Quantum Walkers Propagation.}} The initialized quantum walker distributions are then propagated using the standard structure of FCIQMC dynamics, as in Eq.~\eqref{eq:probs_walkers}. It is worth highlighting here that the diffusion process $\text{QMC}^{(m)}_{g}$ has $G$ independent configurations and $M$ serial evaluations. The entire pipeline can be rewritten in terms of quantum walkers as:
\begin{equation}\label{eq:qmc_walk}
\begin{split}
    \widetilde{d}^{(m)}_{r}(\tau+\Delta\tau)=&-\Delta\tau(\widetilde{H}_{r}-\mathcal{E}^{(m)})\widetilde{d}^{(m)}_{r} + \\ &
    +\Delta\tau\sum_{s}\widetilde{H}_{rs}\widetilde{d}^{(m)}_{s},
\end{split}
\end{equation}
with $\ket{\Psi^{(m)}(\tau)} = \sum\limits_{r}\widetilde{d}^{(m)}_{r}(\tau)\ket{\widetilde{\psi}_{r}^{(m)}} = \sum\limits_{r}\widetilde{d}^{(m)}_{r}(\tau)U_{g}\ket{b_{r}}$, and crucially $\widetilde{H}_{rs} =\bra{\widetilde{\psi}_{r}^{(m)}} H \ket{\widetilde{\psi}_{s}^{(m)}}$. The terms in Eq.~\eqref{eq:qmc_walk} can be readily interpreted, in analogy with the classical FCIQMC, as:
\begin{itemize}
    \item the probability of spawning a quantum walker $s$ from a walker $r$, $\propto |\widetilde{H}_{rs}|\Delta\tau$ 
    \item the spawned quantum walker will inherit the same phase (or sign) as the parent if $\widetilde{H}_{rs}>0$, otherwise it will get a phase of $-\widetilde{H}_{rs}/|\widetilde{H}_{rs}|$
    \item the probability of cloning (killing) the quantum walker $r$ is $\propto |(\widetilde{H}_{r}-\mathcal{E}^{(m)})|\Delta\tau$ if the quantity inside the absolute value is greater (smaller) than zero
    \item quantum walkers with the same label but opposite signs annihilate each other.
\end{itemize}
As we are dealing with $M$ target states, the pipeline above is enriched by one additional step if $M>1$
\begin{itemize}
    \item  if $m>0$ perform the subspace orthonormalization as in Eq.~\eqref{eq:app_orton} and run pipeline for all $m$ in series.
\end{itemize}
At each time step $\Delta\tau$, after the spawning/(killing)cloning/annihilation, the distributions of the number of quantum walkers obtained, one for each of the $m$ targets, $N^{(m)}_i$ are stored, together with the associated signs, $\text{sign}^{(m)}_{i}$. Thus, a stochastic tomography of the quantum states can be realized, namely:
\begin{equation} \label{eqn:walker_tomography}
    \ket{\widetilde{\Psi}^{(m)}(\tau)}=\frac{1}{\sqrt{\mathcal{N}}}\sum_{i=0}^{\dim(H)}\text{sign}^{(m)}_{i}(\tau)N^{(m)}_{i}(\tau)\ket{\widetilde{\psi}^{(m)}_{i}},
\end{equation}
where $\mathcal{N}=\sum_{i=0}^{\dim(H)}[N^{(m)}_{i}(\tau)]^{2}$. 
In the pipeline above, it is clear that the Hamiltonian matrix elements are fundamental for the quantum walker dynamics, as seen in Eq.~\eqref{eq:probs_walkers}. However, when using a quantum circuit as a state ansatz, these quantities need to be evaluated with a quantum algorithm. We make use of the quantum circuits as depicted in Fig.~\ref{fig:modHij_U} for estimating all the non-vanishing transitional states $\ket{\widetilde{\psi}_{s}^{(m)}}$ for a fixed input state $\ket{\widetilde{\psi}_{r}^{(m)}}$ and those in Fig.~\ref{fig:P_overlap} to estimate the $\text{Re}(\widetilde{H}_{rs})$ between the relevant transitions indices. These quantum circuits are a modified version of the standard Hadamard test~\cite{Faehrmann_2025}, with little structural modifications to those constructed in the work~\cite{qcqmc_tp}.
There are relevant hyperparameters in the QMC diffusion that may crucially affect the dynamics and therefore the convergence of the results. First, we need the total evolution time $T$ and the time step $\Delta \tau$, which sets the number of iterations of a single Monte Carlo run, or \emph{trajectory}. The initial number of walkers and the number of independent QMC realizations, which we call trajectories, establish a trade-off: one could run more trajectories with a smaller initial number of walkers in order to obtain similar results by computing the mean of the different trajectories. An initial shift energy $\mathcal{E}^{(m)}(0)$ is provided, which usually is the energy of the initial state used in the preparation stage, and it is dynamically updated in the next iterations as 
\begin{equation}
    \mathcal{E}^{(m)}(\tau + \Delta \tau) = \mathcal{E}^{(m)}(\tau)-\frac{\varepsilon}{\Delta \tau}\log\left(\frac{N^{(m)}(\tau+\Delta\tau)}{N^{(m)}(\tau)}\right),
\end{equation} 
with $N^{(m)}(\tau)=\sum_{i}N_{i}^{(m)}(\tau)$ is the total number of quantum walkers at time $\tau$, and $\varepsilon$ is a damping parameter for further control of the population dynamics. Usually, this is of the same order of magnitude as $\Delta t$. The generation of the quantum walker population is stochastic and may saturate, explode, or disappear. For this reason, it is crucial to tune these hyperparameters to maintain a balance between a substantial number of walkers (sampling efficiently) and the computational cost (memory storage).
\begin{algorithm}
\caption{Cross-domain QCQMC algorithm}~\label{alg:quantum_proc}
\begin{algorithmic}

\Statex \textbf{// Step 1: Define $\textbf{f}(H)$}\\
Encode $f(H)$ in Hamiltonian operator, $H$.\\
Define starting basis $|b_{i}\rangle$
\If{$H$ possesses known symmetries}
    \State $H \gets \text{SymmetryReduction}(H)$
\EndIf \\
Define diffusion strategies $\text{QMC}^{(m)}_{g}$: $m$  serial and $g$ independent evaluations. 
\For{each $g$}

\Statex \textbf{// Step 2: Determine $U_{g}$}
\For{$m=0\dots M-1$}
\State Define $\ket{\widetilde{\psi}_i^{(m)}} = U_{g} \ket{b_i}$

\Statex \textbf{// Step 3: Expectation Value \& Sparsity Check}
\State Compute $|H^{(m)}_{rs}| = |\langle\widetilde{\psi}_r^{(m)} | U_{g}^\dagger H U_{g} | \widetilde{\psi}_s^{(m)}\rangle|$ circuits
\State Identify set of states $s^* = \{s \mid |H^{(m)}_{rs}| \neq 0\}$

\Statex \textbf{// Step 4: Execution of a $\text{QMC}^{(m)}_g$}
\For{each $\tau_i$ in a range of $\tau = 0, \dots, T / \Delta \tau$}
\For{each $s \in s^*$}
    \State Spawning-Death/Clone according to $H_{rs}$
    \State Execute sign circuits for obtaining $H_{rs}$
    \State Annihilate opposite quantum walkers
    \If{$m<M$}
    \State $\ket{\widetilde{\psi_i}^{(m+1)}} \gets \text{$m$-Orthonormalize}(\ket{\widetilde{\psi_i}^{(m)}})$
\EndIf \\

\EndFor
    \State Estimate target quantities with $\overline{f^{(m)}}(H)$
\EndFor 
\Statex \textbf{// Step 5: Store all quantities of interest}
\EndFor
\Statex \textbf{// Step 6: Average $\overline{f^{(m)}}(H)$ over $g$}
\EndFor

\Statex \textbf{// Step 7: Restart the pipeline for several trajectories if needed}
\State $H_{rs}$ should be stored in cache for better performance

\end{algorithmic}
\end{algorithm}

\paragraph{\textbf{Target quantities estimation.}}
Once the $\text{QMC}^{(m)}_{g}$ diffusions are initiated, we calculate, for each time step $\Delta\tau$, a statistical estimator of the target quantities $\overline{f^{(m)}}(H)$. A large family of possible estimators can be found in the literature. Their complete review falls beyond the scope of this work. For the nature of most of the situation we are considering, the most obvious set of estimators are energies or spectral quantities.
For the ground state energy estimation, $m=0$ and $|\widetilde{\Psi}_{0}^{(0)}\rangle$, the most resilient estimator form is the so-called \emph{projective energy}:
\begin{equation}
    \label{eq:en_est_proj}
    E^{0}_{\rm{pr}}(\tau)=\frac{\bra{\widetilde{\psi}_{0}^{(0)}}H\ket{\widetilde{\Psi}^{(0)}(\tau)}}{\langle\widetilde{\psi}_{0}^{(0)}| \widetilde{\Psi}^{(0)}(\tau)\rangle},
\end{equation}
where the $\ket{\widetilde{\psi}_{0}^{(0)}}=U_{g}\ket{b_{0}^{(0)}}$ is the initial ground state guess in the transformed basis. Eq.~\eqref{eq:en_est_proj} tends to the exact $E_{GS}$ when $\tau \rightarrow \infty$.
Another possible estimator, generally more prone to bias than the projective, while less sensitive to the overlap with the initial guess, is the standard variational energy:
\begin{equation}
    \label{eq:en_est_var}
    E^{(m)}_{\rm{var}}(\tau)=\frac{\bra{\widetilde{\Psi}^{(m)}(\tau)}H\ket{\widetilde{\Psi}^{(m)}(\tau)}}{\langle\widetilde{\Psi}^{(m)}(\tau)| \widetilde{\Psi}^{(m)}(\tau)\rangle},
\end{equation}
which need bias regularization via the replica trick~\cite{Brand2022}. Notice that in this form, this quantity guarantees to obtain the exact $E^{(m)}$ when $\tau \rightarrow \infty$. 
Lastly, the estimator may need to take into account for the bias introduced by the shift $\mathcal{E}^{m}$~\cite{GhanemAlav}:  
\begin{equation}\label{eq:en_est_var_sf}
    E^{(m)}_{\rm{sf}}(\tau)=\frac{\sum_{t\le\tau}w_{t}\langle\widetilde{\Psi}^{(m)}(t)| H|\widetilde{\Psi}^{(m)}(t)\rangle}{\sum_{t\le\tau}w_{t}\langle\widetilde{\Psi}^{(m)}(t)| \widetilde{\Psi}^{(m)}(t)\rangle},
\end{equation}
where $w(t)=\prod_{s=1}^{t}{e^{-(\mathcal{E}^{(m)}(s)-D)\Delta\tau}}$ and $D$ is a renormalization factor. This correction to the energy, as weights and accumulates statistics over time, allows us to infer the properties of each time-slice (accumulated in each $\tau$), with reduced bias, hence it is the best choice when the quantity of interest is not found asymptotically. 
It is worth noting that the estimators above do not constitute an exhaustive list of strategies; however, they are sufficient for the simulations described in the following sections.
\subsection{Generation of the state preparation unitary, \texorpdfstring{$U_{g}$}{U g}}
\label{sec:unitary}
The crucial point of the generalized pipeline above stands on the selection of the problem-dependent basis preparation protocol $U_{g}$. In the following, we describe all the strategies investigated in this work. Although these are not a complete set of possibilities, they constitute, in our view, a robust proof of concept of the pipeline. The majority of the $U_{g}$ are here constructed leveraging various forms of variational quantum algorithms: when these are employed, the underlying a set of parametrized quantum circuits is employed. The structures of these ansatzes strongly affect the expressivity of the new basis selected and the scaling and symmetry properties of the problem, hence they need to be designed carefully.
In this work, we utilize a diverse set of reference ansatz, including the well-established Unitary Coupled Cluster Singles and Doubles (UCCSD), the specific Symmetry-Preserving Ansatz (SPA), and a tailored Hardware Efficient Ansatz (HEA); their respective methodologies and applications are detailed forthwith. The UCCSD ansatz~\cite{anand2022quantum} has the form
\begin{equation}\label{eq:uccsd_ansatz}
    \ket{\Psi(\bm{\theta})} = e^{\Xi(\bm{\theta}) - \Xi(\bm{\theta})^{\dagger}} \ket{\bm{b}_0}, 
\end{equation}
where $|\bm{b}_0\rangle$ is a reference state (usually a single configuration of the system) and $\Xi = \Xi_1 + \Xi_2 + \ldots + \Xi_n$ is the $n^{\rm{th}}$ body order excitation terms, that is 
\begin{equation} \label{eq:uccsd_ansatz_terms}
    \begin{split}
        \Xi_1(\theta_{ia}) &= \sum_{ia} \theta_{ia} c^{\dagger}_a c_i, \\ 
        \Xi_2(\theta_{ijab}) &= \sum_{ijab} \theta_{ijab} c^{\dagger}_a c^{\dagger}_b c_j c_i,
    \end{split}
\end{equation}
for first and second order of excitations, respectively, where indices $i,j$ ( $a,b$)  refers to occupied (virtual or unoccupied) orbitals in the reference state and $c^{\dagger}_k$ ($c_k$) are fermionic creation (annihilation) operators over the orbital $k$.
It is convenient, and hence hereby adopted, to consider the first-order Trotter approximation~\cite{trotterI,trotterII}. This effectively allows us to express $U_{\text{UCCSD}}$ as a product of excitations, where each term can be implemented in a quantum circuit form by the means of CX (CNOT) ladders~\cite{cowtan2020generic}.

As an alternative parametrization, we also consider the layered symmetry-preserving ansatz (L-SPA) as introduced in~\cite{gard2020efficient}. Starting again from an initial computational basis state $|b_0\rangle$ the variational state is prepared as
\begin{equation}
\ket{\Psi(\bm{\theta})} =
\prod_{l=1}^{L} U_{\text{SPA}}^{(l)}(\bm{\theta}_l) \ket{b_0} ,
\end{equation}
where $L$ denotes the number of layers and each $U_{\text{SPA}}^{(l)}$ is composed of parametrized two-qubit operations that preserve the chosen symmetry. In this work, we choose to constrain the number of particles in the system, and thus, under a Jordan-Wigner mapping to the Pauli algebra, these gates conserve the Hamming weight of the computational basis states, ensuring that the quantum state remains within the subspace defined by
\begin{equation}
\sum_{i=1}^{N_q} x_i = n_{\text{particles}},
\end{equation}
where $x_i$ is the projection of qubit $i$ and $N_q$ the total number of qubits considered. This ansatz can also be applied to graph-optimization problems, where now the Hamming weight correspond to the number of nodes in a certain graph. Compared to the UCCSD ansatz, which also conserves the Hamming weight via number-conserving excitations, L-SPA presents a notable advantage in terms of a significantly lower circuit depth, albeit potentially leading to a less expressive resulting ansatz.

The Hardware Efficient Ansatz (HEA)~\cite{kandala2017hardware} is specifically designed to leverage the native connectivity and gates available on a particular quantum platform, often with the goal of minimizing circuit depth and the effect of noise. HEAs are typically constructed from alternating layers of single-qubit rotations and entangling gates arranged in a fixed, repeating pattern. A common structure for an $L$-layered HEA can be expressed as
\begin{equation}
    |\Psi(\bm{\theta})\rangle = \prod_{l=1}^L R_y (\theta_y^{(l)}) R_z (\theta^{(l)}_z) U_{\rm{ent}}^{(l)} |b_0\rangle,
\end{equation}
where $U_{\rm{ent}}^{(l)}$ represents a layer of fixed entangling gates (e.g., CX or CZ gates) that connect typically adjacent qubits. While HEAs offer high expressibility and can be readily implemented on current NISQ devices due to their compact structure, their problem-agnostic nature can make them susceptible to barren plateaus~\cite{mcclean2018br} and local minima~\cite{anschuetz_beyond_2022} and, unlike the former ansatz\"{e} considered, they do not generally preserve the symmetries of the system in consideration.

Although the Variational Quantum Eigensolver (VQE)~\cite{tilly2022variational,peruzzo2014variational,mcclean2016} is a natural fit for optimizing the parameters of the UCCSD, SPA, and HEA ansatz\"{e}, our study further investigates alternative variational and non-variational approaches to assess their performance across different optimization strategies.

\subsubsection{Variational Fast Forwarding (VFF)}

As mentioned above, VQE is typically limited to ground state problems, and as such can struggle when higher energy states are targeted. To address this in the generalized QCQMC pipeline, we first introduce the variational fast forwarding (VFF) ansatz as an alternative.
The goal of VFF is to approximate the target unitary operator $U_g$, with a specific product structure $U_{\mathrm{approx}}(\bm{\theta}, \bm{\phi})
= V(\bm{\theta})^{\dagger} \, D(\bm{\phi}) \, V(\bm{\theta})$, where $V$ is a parameterized unitary and $D$ is a parameterized diagonal unitary operator. This variational ``diagonalization'' is accomplished by implementing $U$ on one quantum register and $U_{\textrm{approx}}$ on another, then classically optimizing parameters $\{\bm\theta,\bm\phi\}$ to minimize a loss function that quantifies the discrepancy between the two unitaries using an entanglement fidelity between each pair of qubits in the registers~\cite{cirstoiu2020variational}. This technique is frequently applied to approximate time evolution operators $U=\exp(-iHt)$, particularly when t is small enough that $U$ can be approximated via methods like the Trotter expansion during the training phase. A key advantage of VFF is its ability to ``fast forward'' the resulting approximation: once $V$ and $D$ are learned for small $t$, the operator for a much larger time, $U_M=\exp(-iHMt)$, can be efficiently constructed as $V(\bm\theta)^{\dagger}D(\bm\phi)^MV(\bm\theta) = V(\bm\theta)^{\dagger}D(M\bm\phi)V(\bm\theta)$ without increasing the quantum circuit depth -- hence the name of the algorithm. 

In this work, we specifically utilize VFF to prepare approximations of excited states. Since VFF approximates a diagonalization of $U$, the columns of the learned unitary $V$ naturally form a basis of approximate eigenstates. By preparing these states as $|D_i\rangle = V(\bm\theta_{\textrm{opt}})|b_i\rangle$), we can then use them as an initial basis for Excited State (ES)-QCQMC simulations. Compared to VQE and similar algorithms, which target approximating the ground state only and have no incentive to approximate excited states with accuracy, VFF and other diagonalization algorithms target approximating all eigenstates. Consequently, VFF (and other approximate diagonalisation algorithms) is better suited to prepare suitable initial bases for excited state problems.

\subsubsection{Variational Unitary Matrix Product Operator (VUMPO)} \label{ssec:vumpo}
While the above variational quantum algorithms (VQE and VFF) can in principle learn static and dynamic properties of quantum systems, they are limited in practice by the difficulty of optimizing circuit parameters on a quantum device, including but not limited to problems such as the absence of efficient backpropagation~\cite{abbas_quantum_2023, bowles_backpropagation_2025}, barren plateaus, and local minima~\cite{anschuetz_beyond_2022}.
To circumvent these issues, recently proposed `train-classical, deploy-quantum' or `classical pre-training' approaches offload optimization to a classical device and use the quantum computer only for the final state preparation, or for effective quantum warm-starts. Tensor network quantum circuits exploit this directly: a tensor network is classically optimized and then compiled to a quantum circuit via one-to-one correspondences between tensors and gates. This has proved useful in quantum simulation~\cite{dborin_matrix_2022, dborin_simulating_2022,gover_fully_2025, gibbs2026low} and quantum machine learning~\cite{recio-armengol_train_2026,herrero-gonzalez_born_2025, rudolph_synergistic_2023}. We adopt the same strategy for QCQMC, using tensor network quantum circuits to prepare approximate ground states or approximate diagonalizing unitaries for excited states.
Among the possible tensor network structures~\cite{banuls2023,orus2019,TNreview}, matrix product states (MPS)~\cite{orus2014practical,schollwock2011density} optimized via the density matrix renormalization group (DMRG)~\cite{white1992density,chan2008introduction,stoudenmire2012studying,ganahl2023density}, projected entangled pair states (PEPS), and others, we use the \emph{variational unitary matrix product operator} (VUMPO) ansatz~\cite{pollmann_efficient_2016}, which constructs a matrix product operator (MPO) from a brickwork circuit of two-qubit unitaries. The VUMPO diagonalization is efficient when the target Hamiltonian is fully many-body localized and hence admits a low bond-dimension representation: for an \(n\)-qubit system, the full set of \(2^n\) eigenstates can be captured by \(\mathcal{O}(n)\) local unitaries.

A general MPO on \(n\) sites takes the form
\begin{equation} \label{eq:general_mpo}
    O = \sum\limits_{\substack{\{\boldsymbol{\gamma}\}\\ \{\boldsymbol{\sigma}\}, \{\boldsymbol{\tau}\}}} A_{\gamma_0 \gamma_1}^{[1]\sigma_1 \tau_1}\dots A_{\gamma_{n-1}\gamma_n}^{[n]\sigma_n \tau_n} \ketbra{\sigma_1 \dots \sigma_n}{\tau_1 \dots \tau_n}
\end{equation}
where \(\sigma_i, \tau_i \in \{0, 1\}\) for local dimension \(d=2\), and \(\gamma_i \in \{1, \dots, D\}\) are bond indices connecting adjacent sites. Rather than parameterizing the \(A\) tensors arbitrarily, Ref.~\cite{pollmann_efficient_2016} uses \(L\) layers of two-qubit unitary gates to build a brickwork circuit that maps exactly to an MPO of the form~\eqref{eq:general_mpo} with bond dimension \(D \leq 2^{2L}\), yielding the VUMPO. 

\begin{figure}[t]
\centering
\includegraphics[width=\linewidth]{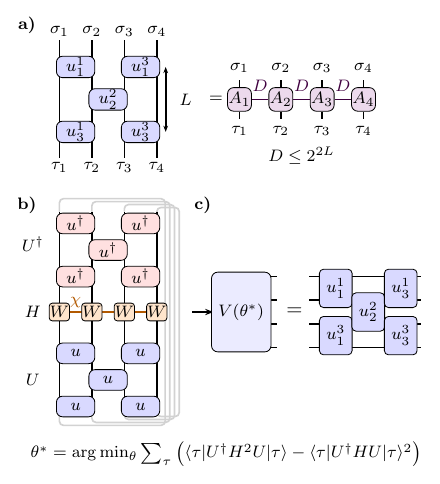}
\caption{Overview of the VUMPO ansatz. (a)~An $L$ layer brickwork network of two-qubit unitaries $\{u^m_\ell\}$ is equivalent to an MPO with bond dimension $D \leq 2^{2L}$. (b) The unitaries are classically optimized using DMRG-like sweeps to approximately diagonalize the Hamiltonian $H$ (written as a bond dimension $\chi$ MPO with tensors \(W\)) by minimizing the total energy variance; looped legs denote contraction over all indices, $\boldsymbol{\tau}$. (c) The optimized brickwork tensor network with parameters $\boldsymbol{\theta}^*$ maps directly to a quantum circuit $V(\boldsymbol{\theta}^*)$ for state preparation targeting ground and/or excited states.}
\label{fig:vumpo}
\end{figure}

In Ref.~\cite{pollmann_efficient_2016} the full spectrum is targeted: the VUMPO tensors are optimized by minimizing a loss function which is the sum of energy variances over all eigenstates (illustrated in Figure~\ref{fig:vumpo} (b)), at a computational cost of \(\mathcal{O}(nD^5\chi^2 d^4)\) per sweep, where $\chi$ is the bond dimension of the target Hamiltonian MPO. Since the resulting brickwork network consists already of two qubit unitaries, it can be directly mapped to a quantum computer. In this way, we bypass the difficulty of performing initial variational eigensolving directly on the quantum device, offloading the expense to a classical device and then address classically intractable components on the quantum device with the QMC pipeline. We give an overview of the VUMPO ansatz in Fig.~\ref{fig:vumpo}.

\paragraph{Hamming-weight symmetry}
The VUMPO can be constrained to enforce Hamming-weight-preserving (HWP) symmetry, which is essential for QCQMC (for example as with the L-SPA VQE ansatz in Section \ref{sec:unitary}). Rather than parameterizing each gate as an arbitrary \(SU(4)\) rotation \(u^k_\ell = e^{iS^k_\ell}\) for site \(m \in \{1, \dots, n\}\) and layer \(\ell \in \{1, \dots, L\}\), where \(S^k_\ell\) is a real symmetric matrix, we restrict to particle-number-conserving gates that act trivially on the \(\ket{00}\) and \(\ket{11}\) sectors and unitarily on the \(\ket{01}, \ket{10}\) subspace:
\begin{equation}\label{eqn:hwp_su4}
    u^k_\ell = \begin{pmatrix}
        1 & 0 &  0 & 0 \\
        0 & (v^k_\ell)_{00} &  (v^k_\ell)_{01} & 0 \\
        0 & (v^k_\ell)_{10} & (v^k_\ell)_{11} & 0 \\
        0 & 0 &  0 & e^{i\phi^k_\ell} \\
    \end{pmatrix}
\end{equation}
where \(v^k_\ell \in \textrm{SU}(2)\) and \(\phi^k_\ell \in [0, 2\pi)\). This restriction confines the VUMPO to a fixed Hamming-weight sector and reduces the number of variational parameters from \(15nL\) (general \(SU(4)\)) to \(4nL\) (three for \(v^k_\ell\) plus \(\phi^k_\ell\)).

\paragraph{The VUMPO quantum circuit} 
After the VUMPO is classically created and optimized, as with VFF we have an approximately diagonalizing unitary, \(V(\boldsymbol{\theta}^*)\) which can be directly implemented as a quantum circuit as in Figure~\ref{fig:vumpo}(c). Here, the parameters \(\boldsymbol{\theta}\) are the elements of the brickwork unitaries, \(\boldsymbol{\theta} = \{e^{iS^k_\ell}\}^k_\ell\) or \(\boldsymbol{\theta} = \{(v^k_\ell)_{ij}, e^{i\phi^k_\ell}\}^k_\ell\) in the Hamming weight preserving case. The VUMPO loss framework is flexible, in the sense that multiple different target quantities e.g. ground $E_{\textrm{GS}}$ or excited $\{E^{(m)}\}_m$ state energies, can be addressed by changing the loss function in the VUMPO classical pre-optimization (i.e., targeting the energy as in a DMRG-like algorithm).

\subsubsection{Haar Unitaries for thermal averages}
The theoretical foundations of FCIQMC stand on the stochastic spawning of Slater determinants and, crucially, on the use of a first order expanded imaginary-time dynamical operator, which serves as a thermalization process for reaching equilibrium (i.e., the ground state energy at low temperature). While the standard procedure for FCIQMC allows us to estimate the ground state energy, and with the aforementioned modifications, the excited state energy as well, it fails to provide an estimate for finite-temperature quantities. This fact emerges from the requirements of the initial state having a non-negligible overlap with the target state and from the fact that FCIQMC is effectively a projection algorithm from the initial state only, leading to a pure state expectation value with no formal equivalence with the trace over the Hilbert space, expected for estimating canonical ensemble quantities. Several strategies have been proposed to circumvent this problem, from SSE~\cite{Yoshioka} to DMQMC~\cite{BluntDMQMC}. In particular, DMQMC is the full density matrix version of FCIQMC, where the typical ``game of life'' played by walkers still takes place, but with walkers now being elements of the density matrix of the system. The idea is as follows: taking a perfectly mixed state in $\text{dim}(H)$-dimensional Hilbert space, $\rho(0)=\frac{\mathbb{I}}{\text{dim}(H)}$, as initial state and propagating it under the symmetrized Bloch equation,
\begin{equation}
    \label{eq:bloch}
    -\frac{\partial \rho}{\partial \beta} = \frac{1}{2}(H\rho + \rho H),
\end{equation}
whose formal solution is $\rho(\beta)=e^{-\beta H/2}\rho(0)e^{-\beta H/2}=e^{-\beta H}/d$, one obtains the correct expression for finite-temperature averages:
\begin{equation}
    \label{eq:therm_av}
    \langle A\rangle_{\beta}=\frac{\text{Tr}(A\rho(\beta))}{\text{Tr}(\rho(\beta))}=\frac{\text{Tr}(A e^{-\beta H})}{\text{Tr}(e^{-\beta H})},
\end{equation}
for any given observable $A$. In DMQMC, the propagation in Eq.~\eqref{eq:bloch} is implemented step-by-step via a first-order (linearized) Euler discretization of the density matrix evolution operator, which preserves Hermiticity at each step thanks to the symmetrized form. Despite its elegance and robustness, DMQMC scales heavily with system size, as it needs to deal directly with density matrices, i.e.\ in the worst case scenario, with a quadratic size expansion of the problem.

We detail here an alternative approach that avoids propagating the density matrix altogether, shifting the cost onto pure state sampling by employing a quantum circuit for basis preparation, inspired by the thermal state perspective developed in Ref.~\cite{Scali_2024_2}. The whole strategy is based on preparing a basis, $U_g$, by employing a unitary Haar operator.
Specifically, we prepare a pure random state $|\xi\rangle=U_{\text{Haar}}|0\rangle^{\otimes n}$, where $U_{\text{Haar}}$ is Haar-distributed~\cite{Plodzien2025}. In the QCQMC implementation, the imaginary-time propagation of $|\xi\rangle$ is realized by expressing the walker dynamics in the full rotated basis $\{U_{\text{Haar}}|b_i\rangle\}_{i=1}^{\text{dim}(H)}$, so that a single QMC diffusion with this basis effectively evolves the initial random state. In this way, averaging over many samples, we obtain the usual equality:
\begin{equation}
    \label{eq:haar_eq}
    \mathbb{E}_{U_{\text{Haar}}}[|\xi\rangle \langle \xi|]=\frac{\mathbb{I}}{\text{dim}(H)},
\end{equation}
effectively reconstructing the perfectly mixed state~\cite{Scali_2024_2}. Now, considering the exact imaginary time evolution operator $e^{-\tau H}$ applied to the random state, we can write the expectation and the normalization factor as, 
\begin{equation}
    \label{eq:single_sample_av}
    \widetilde{\langle A\rangle}=\langle \xi|e^{-\tau H} A e^{-\tau H}|\xi \rangle, \quad \mathcal{N}=\langle \xi|e^{-\tau H}  e^{-\tau H}|\xi \rangle,
\end{equation}
and taking the expectation over different realizations of Haar unitaries, with $\tau=\beta/2$. Inserting the Haar identity~\eqref{eq:haar_eq} and using the cyclicity of the trace, $\mathbb{E}[\langle\xi|M|\xi\rangle]=\text{Tr}[M\,\mathbb{E}[|\xi\rangle\langle\xi|]]=\text{Tr}[M]/d$, we obtain,
\begin{align}
    \label{eq:therm_av2}
    \langle A\rangle_{\beta}=\frac{\mathbb{E}_{U_{\text{Haar}}}[\widetilde{\langle A\rangle}]}{\mathbb{E}_{U_{\text{Haar}}}[\mathcal{N}]}&=\frac{\mathbb{E}_{U_{\text{Haar}}}[\langle \xi|e^{-\tau H} A e^{-\tau H}|\xi \rangle]}{\mathbb{E}_{U_{\text{Haar}}}[\langle \xi|e^{-\tau H}  e^{-\tau H}|\xi \rangle]}\nonumber \\&= \frac{\text{Tr}[Ae^{-2\tau H}]}{\text{Tr}[e^{-2\tau H}]}=\frac{\text{Tr}[Ae^{-\beta H}]}{\text{Tr}[e^{-\beta H}]},
\end{align}
which is exactly Eq.~\eqref{eq:therm_av}. We stress that this derivation holds for the exact propagator $e^{-\tau H}$; in practice, QCQMC approximates it step-by-step via the linearization $e^{-\Delta\tau H}\approx\mathbb{I}-\Delta\tau H$, incurring an error of order $O(\Delta\tau^2)$ per step. It is also important to stress that the thermal average is recovered as the ratio of two independently averaged quantities: the numerator $\mathbb{E}_{U_{\text{Haar}}}[\widetilde{\langle A\rangle}]$ and the denominator $\mathbb{E}_{U_{\text{haar}}}[\mathcal{N}]$ must each be accumulated separately across the Haar samples before dividing, rather than averaging the per-sample ratios $\widetilde{\langle A\rangle}/\mathcal{N}$, which would introduce a bias of order $O(1/\sqrt{G})$ where $G$ is the number of Haar realizations.

Hence, provided a $\Delta \tau$ sufficiently small to support the linearized version of the imaginary time evolution operator, a sufficient sample size of Haar unitaries, and averaging as in Eq.~\eqref{eq:therm_av2}, QCQMC allows the estimation of finite-temperature quantities from pure-state dynamics. This is a remarkable feature in principle, but it inherits the same structural limitations of QCQMC for estimating ground/excited energies (sign problem, basis dependence, and walker-population fluctuations). Furthermore, realizing a full Haar unitary on a quantum circuit is expensive as the system size grows\cite{Haferkamp2022}. However, this problem can be circumvented by employing a unitary $t$-design, which reproduces the moments of the Haar distribution up to order $t$ \cite{Mele2024}. Because only the second moment of the state distribution enters the derivation (Eq.~\eqref{eq:haar_eq}), a $2$-design suffices, with a circuit depth of order $\mathcal{O}(n)$, where $n$ is the number of qubits employed \cite{Harrow2023}.

\subsection{Hamiltonian models, $H$ for $f(H)$}\label{sec:hamiltonians}

We now provide more details about the specific Hamiltonian representations, $H$ for each problem $f(H)$, used as benchmark in this work.

\subsubsection{Molecular chemistry }

The behavior of electrons within a molecule, for a fixed nuclear geometry, is governed by the electronic structure Hamiltonian~\cite{helgaker2013molecular, szabo2012modern}. This Hamiltonian arises from the Born-Oppenheimer approximation, which assumes that the much heavier nuclei remain stationary relative to the rapidly moving electrons. Under this approximation, the electrons move in the electrostatic field created by the nuclei, and the total electronic energy for a given nuclear configuration can be determined.

For a system of electrons, the electronic structure Hamiltonian is commonly expressed in a second-quantized form. This formalism provides a compact and systematic way to represent many‑body interactions while enforcing the fermionic antisymmetry of the electronic wavefunction. In an orthonormal basis of molecular spin-orbitals, it is written as:
\begin{equation}\label{eq:sq_hamiltonian}
H_{\rm{elec}} = \sum_{\alpha \beta} h_{\alpha \beta} a^\dagger_\alpha a_\beta + \frac{1}{2}\sum_{\alpha \beta \gamma \delta} h_{\alpha \beta \gamma \delta} a^\dagger_\alpha a^\dagger_\beta a_\gamma a_\delta,
\end{equation}
where $a^{\dagger}_\alpha$ and $a_\alpha$ are the creation and annihilation operators, respectively, for an electron in the spin-orbital $\alpha$. 
The terms $h_{\alpha \beta}=-\langle \alpha|\frac{1}{2}\nabla^2+\sum_A \frac{Z_A}{r_A} |\beta\rangle$ represents the one-electron integrals, which take into account the kinetic energy of the electrons and their attractive Coulomb interaction with the nuclei (with charge $Z_A$ at position $r_A$). Conversely, the $h_{\alpha \beta \gamma \delta} = \langle \alpha \beta| r_{12}^{-1} |\gamma \delta\rangle$ are the pair-interaction integrals, which describe the repulsive Coulomb interaction between pairs of electrons (where $r_{12}$ is the distance between electron 1 and electron 2). The indices $\alpha, \beta, \gamma, \delta$ refer to specific spin-orbitals. 

In principle, one could obtain the full spectrum by constructing the Hamiltonian matrix in a Slater‑determinant basis and diagonalizing it. However, the size of this matrix grows exponentially with the number of spin‑orbitals and the exact computation of the spectrum becomes intractable already when a few dozen of electrons are considered.  
A common strategy to mitigate this exponential scaling is to restrict the calculation to an active space, in which only a selected subset of electrons and orbitals (the active electrons and active orbitals) are treated explicitly. While this truncation introduces an approximation, it significantly reduces the effective Hilbert‑space dimension and often captures the chemically relevant correlations. Applying the Jordan–Wigner transformation~\cite{JordanWigner} maps the fermionic creation and annihilation operators in Eq.~\eqref{eq:sq_hamiltonian} to qubit operators. The resulting qubit Hamiltonian is expressed as a weighted sum of Pauli strings, as in Eq.~\eqref{eq:lcu_ham}, enabling its use in quantum phase estimation or variational quantum algorithms in a quantum computer.

\subsubsection{2D Fermi-Hubbard model}
\begin{figure}
\includegraphics[width=\columnwidth]{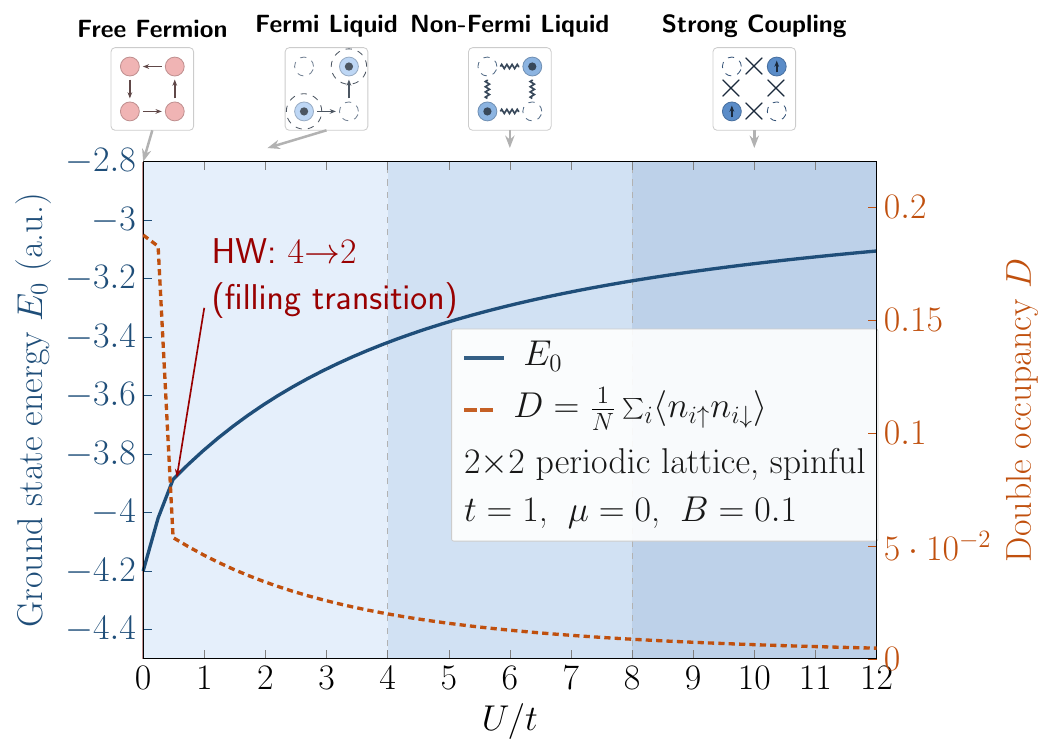}
    \caption{
    pPhase diagram of the \(2\times 2\) Fermi-Hubbard model with \(\mu=0\) showing the free-Fermion (\(U=0\)), Fermi liquid (\(U/t \lesssim 4\)), Non-Fermi liquid (\(4\lesssim U/t \lesssim 8\)) and strong coupling (\(U/t \gg 8\)) phases. For half filling, \(\mu = \frac{U}{2}\), the strong coupling phase becomes a Mott insulating phase.
    }
    \label{fig:fh_phase}
\end{figure}

One of the cornerstone models of electronic systems and materials is the Fermi-Hubbard model. Despite its simplicity, it shows a wide variety of phenomena, even in its one-dimensional formulation~\cite{arute2020observation}. In two dimensions, the Fermi-Hubbard Hamiltonian is considered a non-trivial benchmark for studying strongly correlated systems and phase transitions. The two-dimensional Fermi-Hubbard Hamiltonian~\cite{2dfh},
\begin{equation}
H_{\text{FH}} = -t \sum_{\langle ij \rangle, \sigma} (a^\dagger_{i\sigma} a_{j\sigma} + a^\dagger_{j\sigma} a_{i\sigma}) + U \sum_i n_{i\uparrow} n_{i\downarrow} - \mu \sum_{i, \sigma} n_{i\sigma},
\end{equation}
represents fermions with spin up or down on a square lattice, with a nearest neighbor hopping term $t$, on-site interaction strength $U$, where $n = a^{\dagger} a$ and chemical potential $\mu$. When spins degrees of freedom are included, which we denote by $\sigma = \{\uparrow, \downarrow\}$, it is possible to explore the rich phase diagram of the model and study the formation of various phases entailed in the equilibrium wavefunction, even for finite systems. At zero chemical potential, tuning the ratio $U/t$ allows us to explore the Fermi- and non-Fermi-liquid phases crossover phase regimes as described in Fig.~\ref{fig:fh_phase}.
As in the preceding section on molecular-chemistry Hamiltonians, exact solutions for finite Fermi-Hubbard systems can be obtained by constructing and diagonalizing their matrix representations.

\subsubsection{Nuclear shell model}

\begin{figure}
    \centering
    \includegraphics[width=0.9\columnwidth]{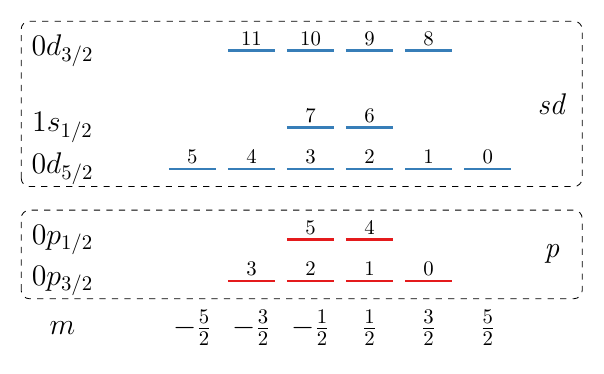}
    \caption{Configurational valence space for the $p-$ and $sd-$shells. The valence space is decomposed into two degenerate $j$ orbitals with degeneracy $(2j+1)$. The labels on the left correspond to the standard spectroscopic notation $n\ell_j$, where $\ell=p$ means $\ell=1$ and $\ell=s,d$ means $\ell=0,2$, respectively. Numbers on top of every single-particle state represent the labels used in this work. Slightly modified from~\cite{perez2023nuclear}.}
    \label{fig:nsm_configuration_space}
\end{figure}

In the nuclear shell model, atomic nuclei are described similarly to how molecules are described in quantum chemistry. This involves treating protons and neutrons (nucleons) as fundamental fermions that occupy single-particle states, or orbitals, within a valence space as shown in Fig.~\ref{fig:nsm_configuration_space}. These orbitals are characterized by quantum numbers: principal quantum number ($n$), orbital angular momentum ($l$), total angular momentum ($j$), its projection ($m$), and isospin's third component ($t_z$), which distinguishes protons from neutrons. 
In the nuclear shell model, the description of a nucleus begins with a second-quantized Hamiltonian similar as to the one described in Eq.~\eqref{eq:sq_hamiltonian}. Its parameters—single-particle energies and two-body interactions—are empirically tuned to reproduce experimental nuclear data. Nuclear states are then described by their total angular momentum ($J$) and isospin ($T$), which are derived from the individual angular momenta ($j$) and isospins ($t_z$) of the constituent nucleons. The third-component projections ($M$ and $T_z$) for the entire nucleus are simply the sum of these projections for its individual nucleons in the valence space.

This allows us to represent a nuclear state with specific $J$ and $T$ values in the many-body basis of nucleons occupying the single-particle orbitals of the specific configurational space, denoted by $M$ and $T_z$:
\begin{equation}
    \ket{JMTT_z} = \sum_{i} c_{i}\ket{i, MT_z},
\end{equation}
where $i$ denotes additional quantum numbers and $c_i$ are the coefficients of the expansion. Nuclear states are ultimately found by solving the eigenvalue problem for the Hamiltonian matrix constructed in this basis. The complex nature of this many-body basis, which accounts for both protons and neutrons, results in a substantial size given by $\binom{ d_\text{sh} }{ N_\text{val} } \binom{ d_\text{sh} }{ Z_\text{val} }$, where $d_\text{sh}$ is the dimension of the valence space, and $N_\text{val}$ and $Z_\text{val}$ are the numbers of valence neutrons and protons.

Quantum algorithms have been widely used to investigate nuclear structure, focusing on both ground-~\cite{kiss2022quantum,dumitrescu2018cloud,romero2022solving,perez2023nuclear,stetcu2022variational,sarma2023prediction,carrasco2025comparison,sarma2026low,costa2025quantum,gibbs2026low} and low-lying excited states~\cite{zhang2025excited,bhoy2024shell,ruiz2022accessing,robin2023quantum}. The FCIQMC method has also recently found application in nuclear structure calculations~\cite{jin2025full}. Additionally, the structure and entanglement properties of nuclear states have also been thoroughly studied~\cite{robin2021entanglement,hengstenberg2023multi,perez2023quantum,tichai2023combining,perez2024entropy}.

\subsubsection{Cardinality-Constrained MaxCut Optimization}
\label{sec_hams_opt}

Optimization problems defined on graphs play a central role in combinatorial optimization, where the objective is to make decisions based on the structure and weights of vertices and edges. The cost function in Eq.~\eqref{eq:maxcut_cardinality} defines a \emph{cardinality-constrained MaxCut problem} on a graph \(G = (V, E)\), where \(V\) is the set of \(n_\mathrm{nodes}\) vertices and \(E\) is the set of edges. This problem generalizes the classical MaxCut by introducing an additional constraint into the objective function. Specifically, consider a weighted graph in which each edge \((i,j) \in E\) has an associated weight \(w_{ij}\). The target of the optimization problem is to find a candidate binary vector \(x = (x_1, \ldots, x_{n_\mathrm{nodes}})\), with \(x_i \in \{0,1\}\), which maximizes the number of weighted edges that cross a subset of the graph  \(\Delta\). The existence of such a subset induces an effective partition of $G$ in \(\Delta\) and \(\bar{\Delta}=G-\Delta\). The \textit{cardinality constraint} enforces that exactly \(n_\mathrm{ones}\) vertices are assigned to the first subset, representing scenarios where only a fixed number of nodes may participate in the cut. In essence, the $f(H)$ for the \textit{cardinality-constrained MaxCut problem} can be formulated as:

\begin{equation}
\begin{aligned}
\text{maximize} \quad & C(x) = \sum_{(i,j)\in E} w_{ij} \, (x_i \oplus x_j) \\
\text{subject to} \quad & \sum_{i=1}^{n_\mathrm{nodes}} x_i = n_\mathrm{ones}, \\
& x_i \in \{0,1\}, \quad i = 1, \ldots, n_\mathrm{nodes},
\end{aligned}
\label{eq:maxcut_cardinality}
\end{equation}
where \(\oplus\) denotes the binary XOR operation.
The corresponding cost Hamiltonian, $H$, consists of two components: the MaxCut Hamiltonian \(H_\mathrm{MaxCut}\), and a penalty term \(H_\mathrm{c}\) enforcing the cardinality constraint, scaled by a parameter \(\lambda\):
\begin{equation}
\begin{aligned}
H &= H_\mathrm{MaxCut} + \lambda \, H_\mathrm{c}, \\
\text{with} \quad
H_\mathrm{MaxCut} &= \sum_{(i,j)\in E} w_{ij} \, Z_i Z_j, \\
H_\mathrm{c} &= \left( \sum_{i=1}^{n_\mathrm{nodes}} \frac{1 - Z_i}{2} - n_\mathrm{ones} \right)^2.
\end{aligned}
\label{eq:hamiltonian}
\end{equation}
Individual vertices are encoded in single qubits on the computational basis state, with $|0\rangle \in \Delta$ and $|1\rangle \in \bar{\Delta}$, respectively. For example, the bitstring \((0,1,1,0)\) corresponds to a 4-node instance in which vertices 1 and 4 belong to set \(\Delta\), while vertices 2 and 3 belong to \(\bar{\Delta}\), satisfying the cardinality constraint \(n_\mathrm{ones}=2\).
This formulation naturally extends to other combinatorial optimization problems that impose a fixed ratio or count of zeros and ones in the optimal bitstring, e.g., the \(k\)-vertex cover problem~\cite{tu2022survey}.

\section{Results}\label{sec:results}

\subsection{Ethylene spectrum energy, VQE/VFF/VUMPO comparison}\label{sec:results_ethylene}

\begin{figure}
\centering
\includegraphics[width=0.3\columnwidth]{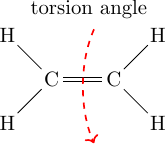}
\caption{Ethylene molecule, composed of two carbon atoms sharing a double bond and four hydrogen atoms. The C-C bond can be twisted by changing the H-C-C-H torsion angle, resulting in planar and perpendicular geometries when the torsion angle is $0^\circ$ and $90^\circ$, respectively.}
\label{fig:ethylene_molecule}
\end{figure}

We first want to estimate the ground and excited states of the ethylene molecule (C$_2$H$_4$) as a function of the torsional angle of the double bond C--C, from $0^\circ$ (planar) to $90^\circ$ (perpendicular), as depicted in Fig.~\ref{fig:ethylene_molecule}. For this study, we restrict the active space of the molecule to two electrons in three orbitals which, under a Jordan-Wigner transformation, translates to quantum circuits and states of six qubits. Using the methods described in the preceding sections, we simulate the molecule for the first five low-lying eigenstates in energetic order as a function of the torsional angle. For $90^\circ$, the ground state of ethylene is a triplet state and thus the first three eigenstates are degenerate in energy. The fourth eigenstate (or first \emph{true} excited state) is also very close in energy. The corresponding details are summarized in Table~\ref{tab:ethylene_exact}. This molecule presents unique challenges for simulation due to these specific characteristics, but the QCQMC method is able to identify and extract the relevant features. 

\begin{table} 
    \centering

    \begin{tabular}{@{} c c c S[table-format=1.4] S[table-format=1.4] @{}} 
        \toprule
        \textbf{Target State} & \textbf{Energy (Ha)} & \textbf{Basis State} & \textbf{Amplitude} \\

        \midrule
        Ground state 1 & -76.913 & 100100 & 0.7071 \\
          & & 011000 & 0.7071 \\
        \midrule 
        Ground state 2 & -76.913 & 101000 & 0.7071 \\
         &  & 010100 & 0.7071 \\
        \midrule
        Ground state 3 & -76.913 & 010100 & 0.7071 \\
         & & 101000 & -0.7071  \\
        \midrule
        Excited state 1 & -76.910 & 011000 & -0.7071 \\
         & & 100100 & 0.7071 \\
         \midrule 
        Excited state 2 & -76.793 & 110000 & 0.9999  \\
        \bottomrule
    \end{tabular}
    \caption{Ethylene energies and bitstring expansions results of the eigenstates for a perpendicular geometry ($90^{\circ}$) obtained from exact diagonalization. The ground state is a triple-degenerate eigenstate, and the first excited state is a singlet with a very low energy gap. Only bitstring contributions to the eigenstate larger than $0.1\%$ have been taken into account. }
    \label{tab:ethylene_exact}
\end{table}

\begin{figure}
    \centering
    \includegraphics[width=0.95\columnwidth]{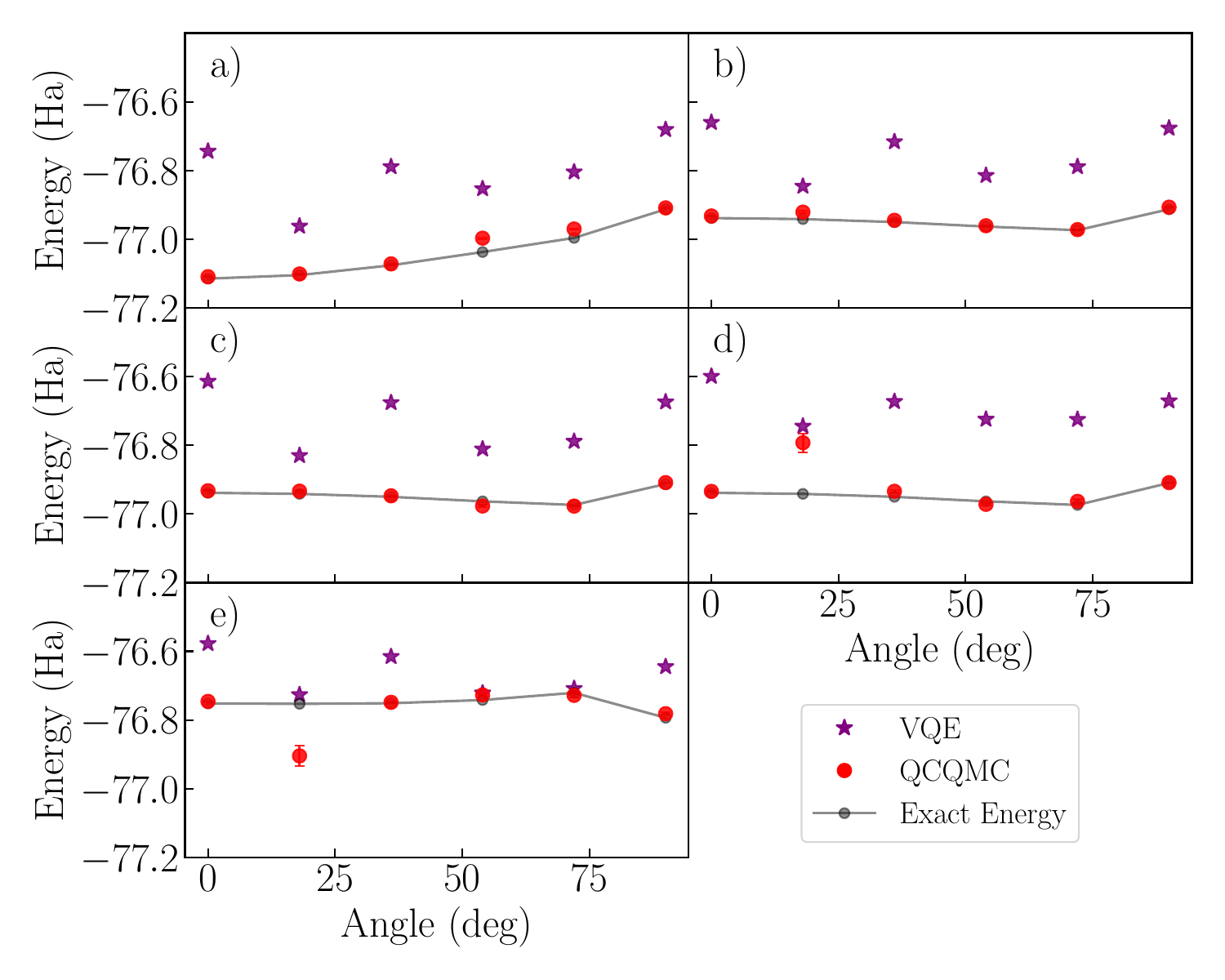}
    \caption{QCQMC results for the final energy, utilizing a VQE-prepared state basis, for the six-qubit ethylene problem as a function of the torsion angle. Figures $a)-e)$ depict the five lowest eigenstates, in corresponding order. Full details of the problem description and chosen hyperparameters are detailed in Sec.~\ref{sec:results_ethylene}. VQE-prepared state energies and eigenenergies obtained with exact diagonalization are shown for benchmarking.}
    \label{fig:ethylene_es_vqe_6q_res}
\end{figure}

\begin{figure}
    \centering
    \includegraphics[width=0.99\columnwidth]{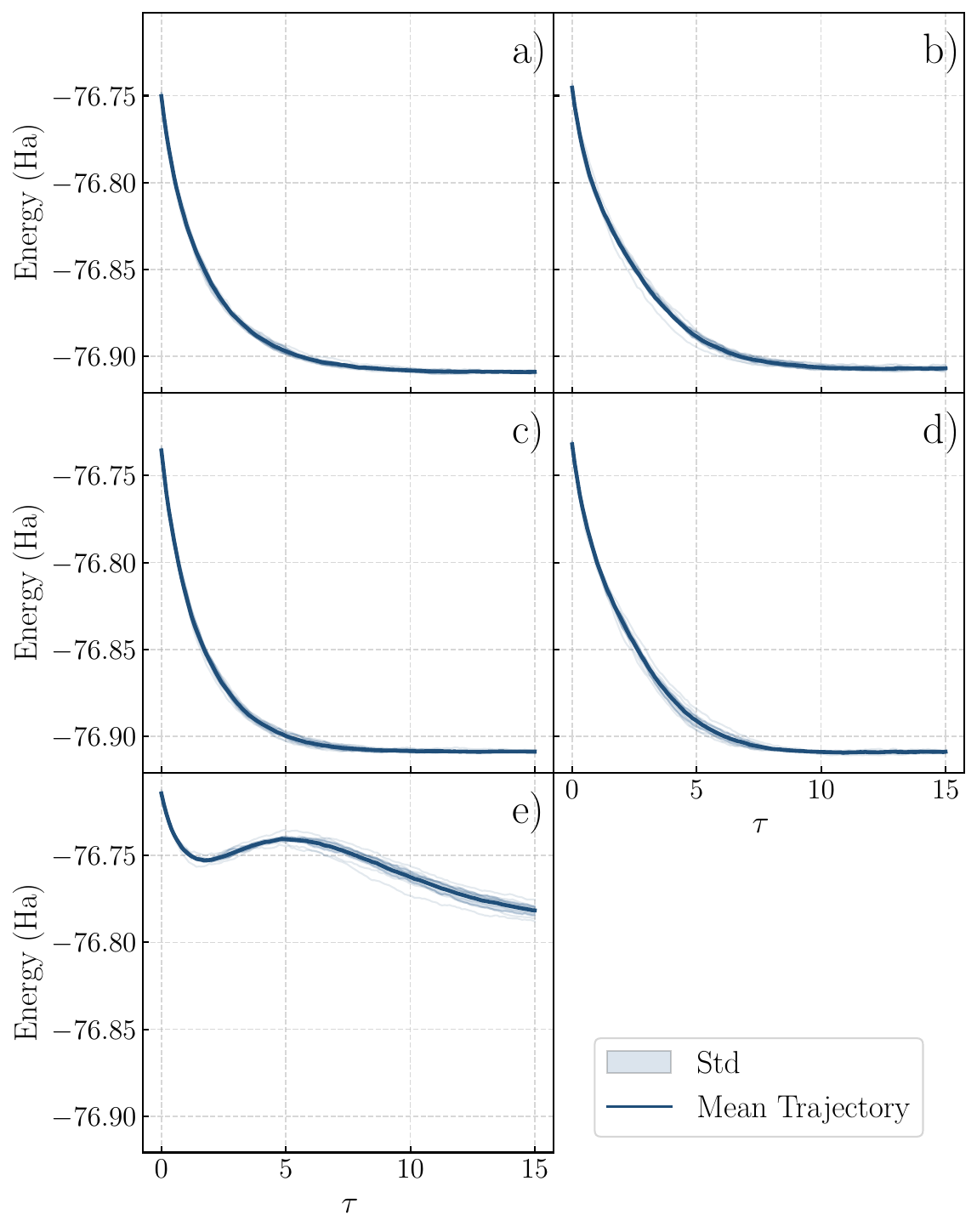}
    \caption{Trajectories of QCQMC with a VQE-prepared state basis, for the six-qubit ethylene problem with perpendicular ($90^{\circ}$) geometry as a function of the time propagation (with units of inverse of energy). Figures $a)-e)$ depict the five lowest eigenstates, in corresponding order. Full details of the problem and hyperparameters are detailed in Sec.~\ref{sec:results_ethylene}. The eigenenergies and further details of the solutions obtained from exact diagonalization are detailed in Table~\ref{tab:ethylene_exact}. The mean of the trajectories is shown as a solid curved line, and the standard deviation as a band surrounding the mean. Notably, the standard deviation remains relatively low across the displayed trajectories.}
    \label{fig:ethylene_es_vqe_6q_traj}
\end{figure}

\begin{figure}
    \centering
    \includegraphics[width=0.95\columnwidth]{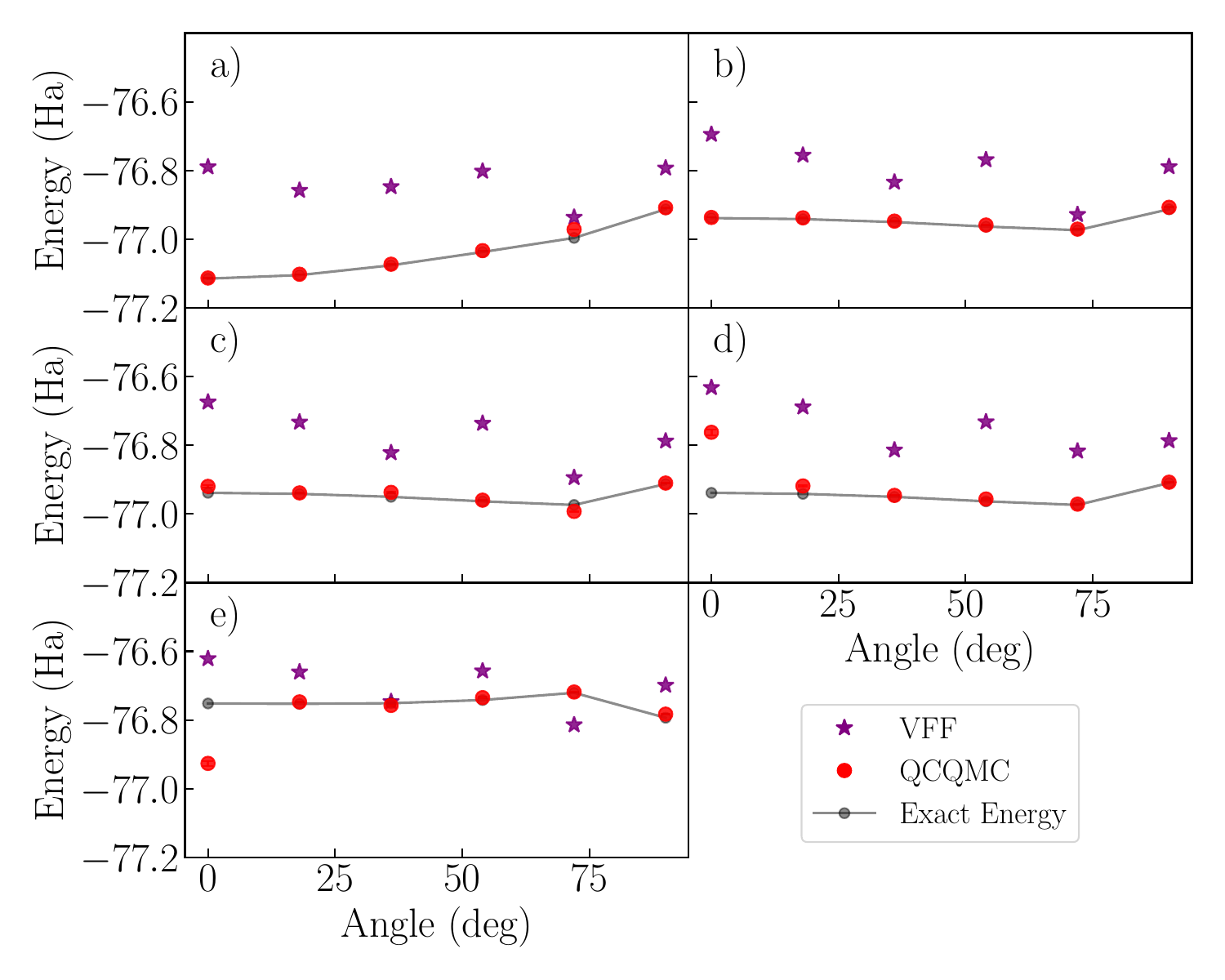}
    \caption{QCQMC results for the final energy, utilizing a VFF-prepared state basis, for the six-qubit ethylene problem as a function of the torsion angle. Figures $a)-e)$ depicts the five lowest eigenstates in order. Full details of the problem description and chosen hyperparameters are detailed in Sec.~\ref{sec:results_ethylene}. VFF-prepared state energies and eigenenergies obtained with exact diagonalization are shown for benchmarking. The lower-bound condition from the variational principle does not extend directly to excited states. Thus, we may observe energies obtained from the state-preparation method that fall below the eigenenergies in these cases.}
    \label{fig:ethylene_es_vff_6q_res}
\end{figure}

\begin{figure}
    \centering
    \includegraphics[width=0.99\columnwidth]{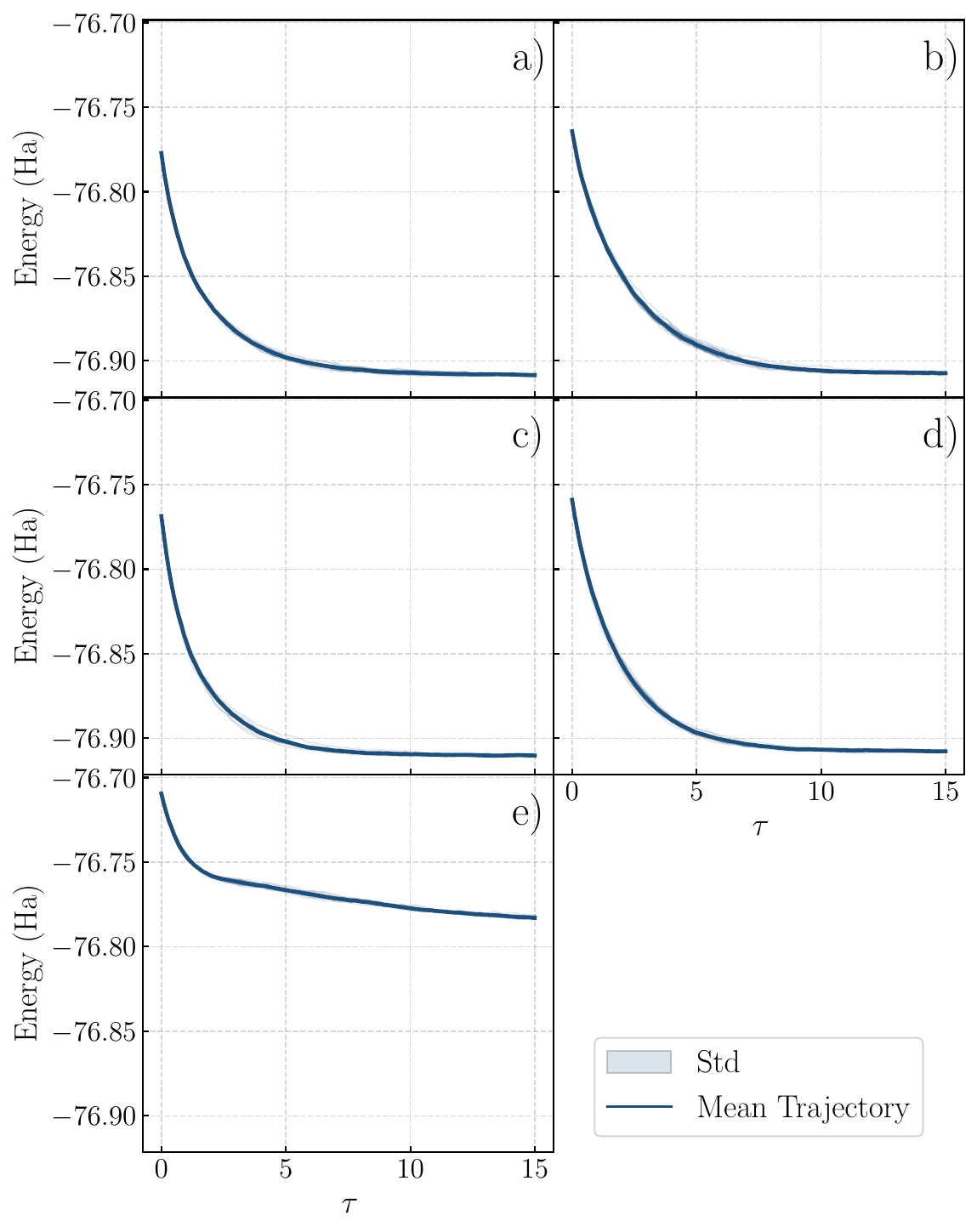}
    \caption{Trajectories of QCQMC with a VFF-prepared state basis, for the six-qubit ethylene problem with perpendicular ($90^{\circ}$) geometry as a function of the time propagation (with units of inverse of energy). Figures $a)-e)$ depict the five lowest eigenstates, in corresponding order. Full details of the problem and hyperparameters are detailed in Sec.~\ref{sec:results_ethylene}. The eigenenergies and further details of the solutions obtained from exact diagonalization are detailed in Table~\ref{tab:ethylene_exact}. The mean of the trajectories is shown as a solid curved line and the standard deviation as a band surrounding the mean. Notably, the standard deviation remains relatively low across the displayed trajectories.}
    \label{fig:ethylene_es_vff_6q_traj}
\end{figure}

First, we compare two different basis preparation techniques for this problem -- VQE and VFF. For both algorithms, we consider the state-preparation unitary $U_g$ as a hardware-efficient ansatz (HEA) over six qubits with unit depth. This is then trained, as the respective algorithm requires, using in both cases an Adam optimizer with a learning rate of 0.1 and 50 maximum iterations. 
QCQMC is then run with $2,000$ initial walkers, a walker threshold of $8,000$, $10$ trajectories, for 150 iterations with a time step of $0.1$. As discussed in Appendix~\ref{sec:fciqmc}, the projector energy estimator~\eqref{eq:en_est} performs poorly for excited states. Therefore, for this excited state problem energy is evaluated as an expectation value. The results using the VQE-prepared basis are shown in Figs.~\ref{fig:ethylene_es_vqe_6q_res} and~\ref{fig:ethylene_es_vqe_6q_traj} and from the VFF prepared basis in Figs.~\ref{fig:ethylene_es_vff_6q_res} and~\ref{fig:ethylene_es_vff_6q_traj}. 

Figures \ref{fig:ethylene_es_vqe_6q_res} and \ref{fig:ethylene_es_vff_6q_res} show and compare the final energy results from the state-preparation technique used and the final corrected result from QCQMC to the solutions from exact diagonalization for all considered torsional angles. Figures~\ref{fig:ethylene_es_vqe_6q_traj} and~\ref{fig:ethylene_es_vff_6q_traj} show the mean and standard deviation of trajectories for QCQMC for the perpendicular geometry of ethylene ($90^\circ$) for the five lowest eigenstates. We see that for this problem, both VQE and VFF perform similarly. Overall, the energy resulting from QCQMC for all angles and energy levels converges closely to the expected energy value, and in particular, is much closer than the energy value provided by the state-preparation methods VQE and VFF.
QCQMC exhibits in both cases a single instance of ``misordering", where QCQMC converges to a different energy level than expected -- e.g., for the VQE example, excited state 1 (ES1) and excited state 2 (ES2) have been swapped for angle $18^\circ$, and for VFF this has occurred at angle $0^\circ$. This generally occurs when the initial basis is such that the initial trial state has much greater overlap with a higher-energy state than with the state it is intended to initialize. Some mixing is also observed between ES1 and the ground states due to the low energy gap between these energies. A potential solution to address state degeneracy and improve QMC performance involves introducing a controlled and parameterized single-body term into the Hamiltonian, which is designed to precisely break the degeneracy while its effect on the exact eigenstates remains known at least on a desired energy window. The trajectories converge closely to the expected energy with low standard deviation, though higher energy states take longer to converge. This is expected because higher-energy states are affected by the convergence of lower-energy states via the projective step. This phenomenon arises from the energetic proximity of the studied eigenstates, which introduces mixing within the QCQMC pipeline. Resolving this issue requires increasing the number of walkers to achieve better energetic resolution.

\begin{figure}
    \centering
    \includegraphics[width=\columnwidth]{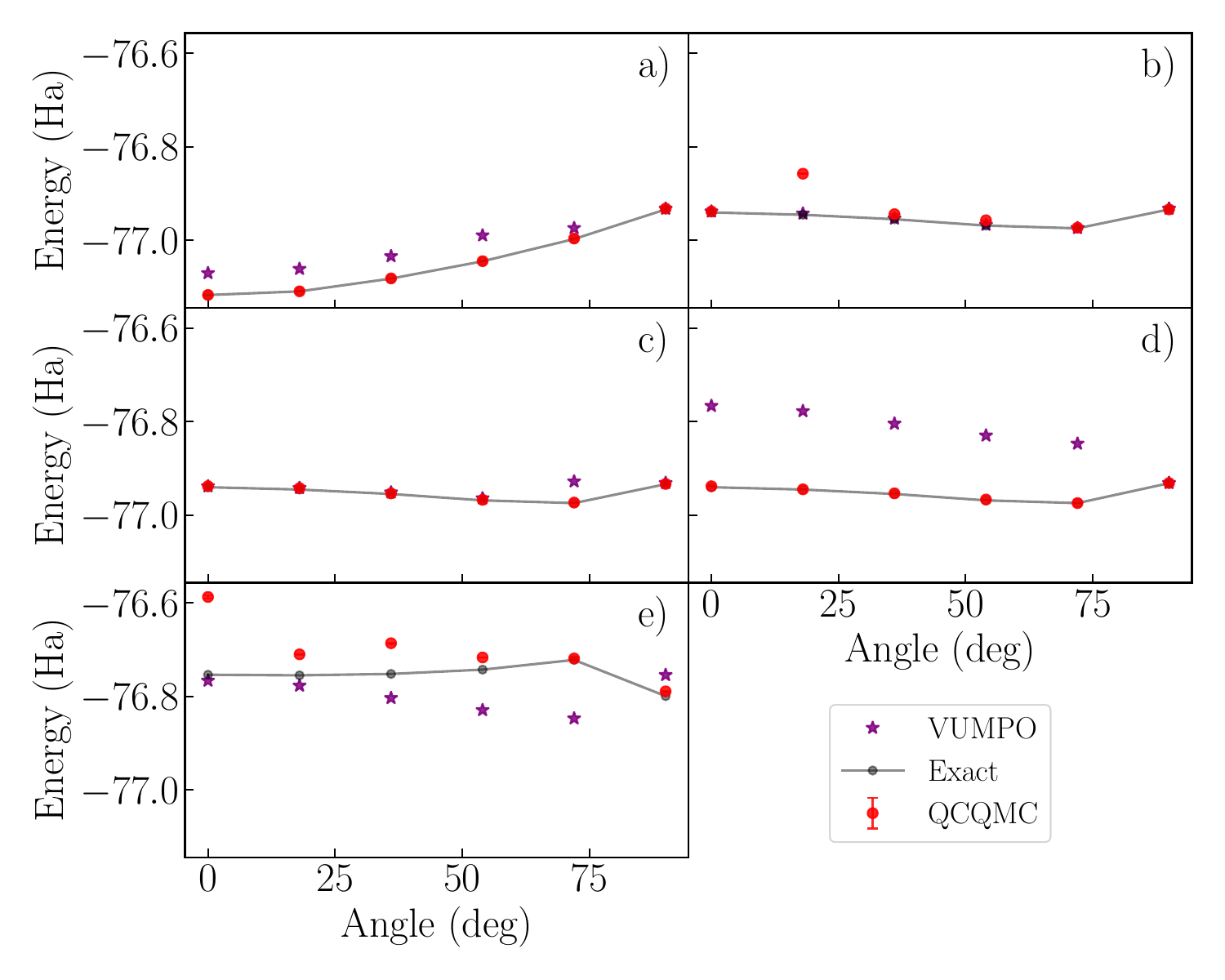}
    \caption{QCQMC results for the final energy, utilizing a VUMPO-prepared state basis, for the eight-qubit ethylene problem as a function of the torsion angle. Figures $a)-e)$ depict the five lowest eigenstates, in corresponding order. Full details of the problem description and chosen hyperparameters are detailed in Sec.~\ref{sec:results_ethylene}. VFF-prepared state energies and eigenenergies obtained with exact diagonalization are shown for benchmarking. The lower-bound condition from the variational principle does not extend directly to excited states, and thus we may observe energies obtained from the state-preparation method that fall below the eigenenergies in these cases.}
    \label{fig:ethylene_es_vumpo_8q}
\end{figure}

\begin{figure}
    \centering
    \includegraphics[width=0.99\columnwidth]{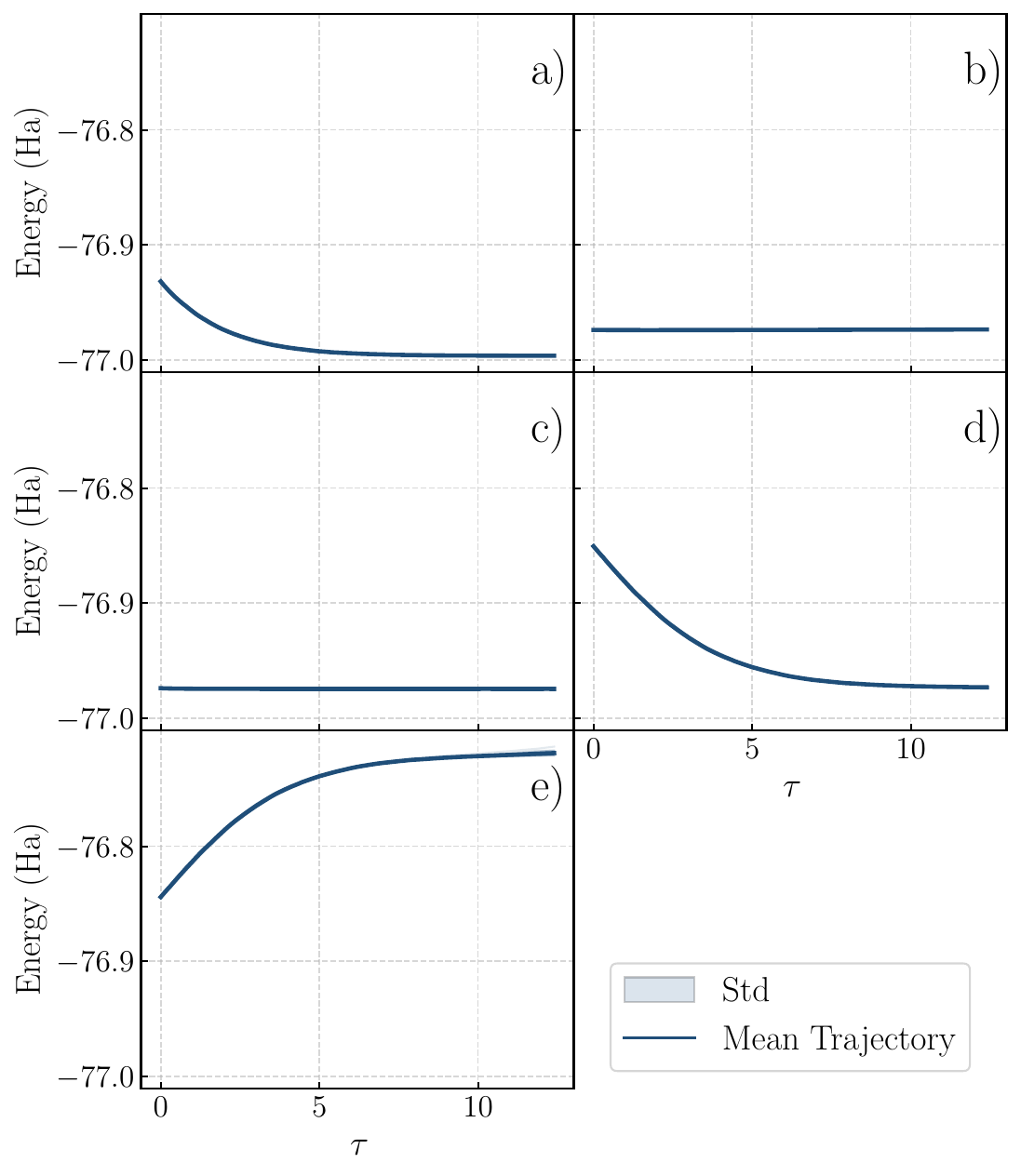}
    \caption{Trajectories of QCQMC with a VUMPO-prepared state basis, for the eight-qubit ethylene problem with perpendicular ($90^{\circ}$) geometry as a function of the time propagation (with units of inverse of energy). Figures $\rm{a})-\rm{e})$ depict the five lowest eigenstates, in corresponding order. Full details of the problem and hyperparameters are detailed in Sec.~\ref{sec:results_ethylene}. The eigenenergies and further details of the solutions obtained from exact diagonalization are detailed in Table~\ref{tab:ethylene_exact}. The mean of the trajectories is shown as a solid curved line and the standard deviation as a band surrounding the mean. Notably, the standard deviation remains relatively low across the displayed trajectories.}
    \label{fig:ethylene_es_vumpo_8q_traj}
\end{figure}

Finally, for completeness, we use VUMPO to scale the problem to a larger instance. Similarly to the results shown for VQE and VFF, the results from using the VUMPO prepared basis are shown in Figs.~\ref{fig:ethylene_es_vumpo_8q} and~\ref{fig:ethylene_es_vumpo_8q_traj} for the ethylene molecule with four electrons in four orbitals, that is, an eight-qubit state problem after a Jordan-Wigner transformation. We use six layers of the brickwork ansatz with a Hamming-weight preserving parameterization for the tensors as per Eq.~\eqref{eqn:hwp_su4}. We still observe (and now correct for) the same misordering behavior as described above for the VUMPO state-preparation method. We use an average over ten trajectories with a time step of $0.1$ (in units of inverse of energy) for $120$ iterations. We also increase the number of initial walkers and threshold to $10,000$ and $15,000$ respectively to ensure convergence. Overall, VUMPO state preparation consistently estimates eigenstate energies accurately. The corrections provided by QCQMC are often negligible, indicating VUMPO's high fidelity in the weakly-correlated regime. 

\subsection{2D Fermi-Hubbard}
Our study investigates the two-dimensional Fermi-Hubbard model on a 2×2 square lattice with periodic boundary conditions at half-filling, which maps to an eight-qubit system for quantum simulation under the Jordan-Wigner transformation. We report the computed ground state properties for both the Fermi liquid (FL) and non-Fermi liquid (NFL) phases, comparing the performance of VQE and VUMPO state preparation approaches. We used a UCCSD ansatz~\eqref{eq:uccsd_ansatz} with two-body excitations for the VQE state-preparation method, where, for the quantum circuit implementation, a single Trotter step and second Trotter order were used. To leverage known symmetries and reduce the variational space, a symmetry-restricted basis was employed. Specifically, the $2\times2$ Fermi-Hubbard Model at half-filling is known to conserve particle number, which corresponds to a fixed total Hamming weight of two in our eight-qubit encoding; this constraint was enforced during state preparation. For the FL and NFL phases, we use an initial number of walkers of $8,000, 5,000$ with a threshold of $10,000, 15,000$ respectively. For the FL phase, we use $8$ VUMPO layers, and for the NFL we use $5$ with $10$ and $8$ iterations respectively.

\begin{figure}
    \includegraphics[width=\columnwidth]{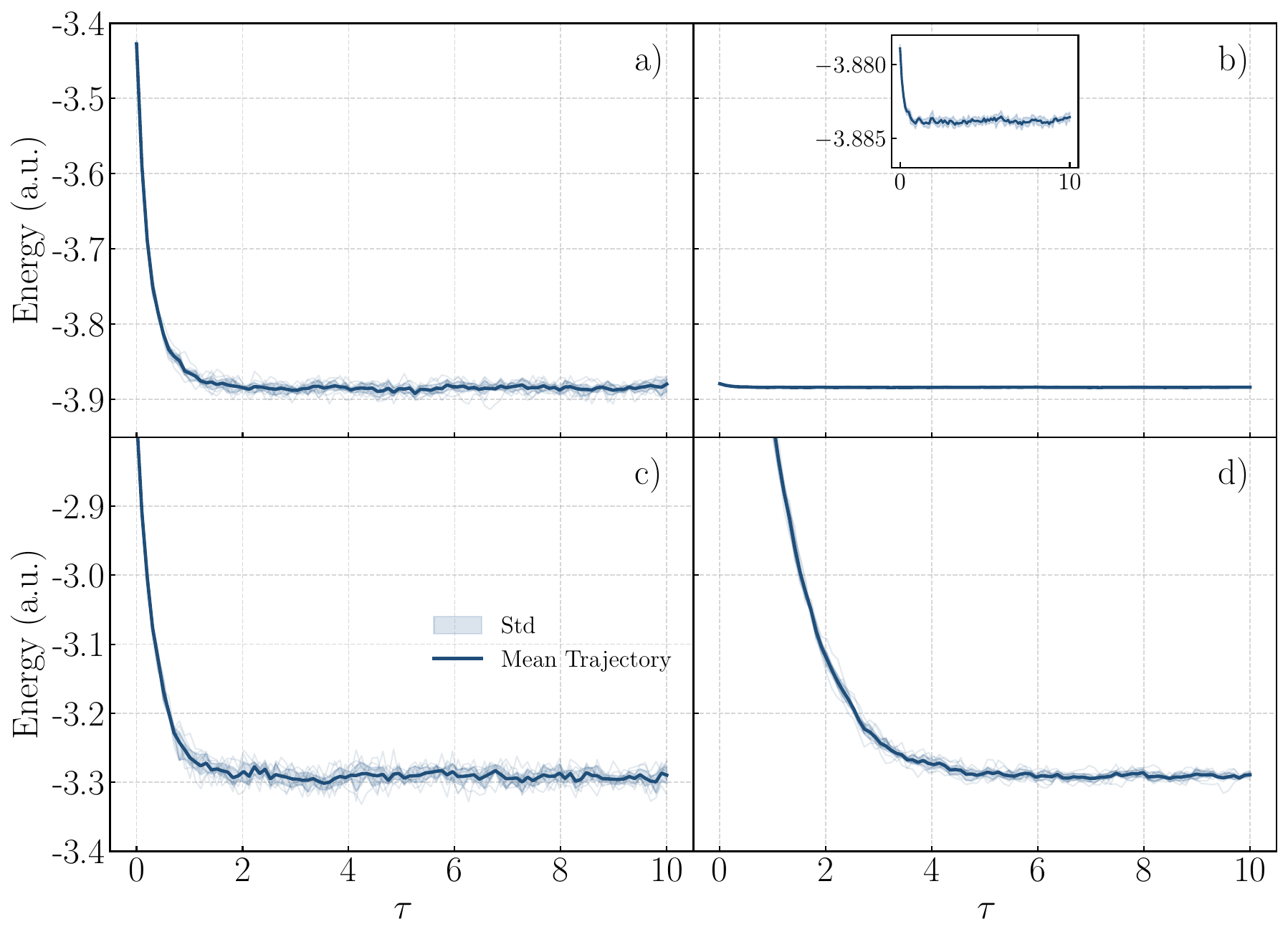}
    \caption{Ground-state trajectory results as a function of the time propagation (with units of inverse of energy) for two phases of the spinful \(2\times 2\) Fermi-Hubbard model with QCQMC: the Fermi liquid ($U/t=0.5$) in Figures $\rm{a})-\rm{b})$ and non-Fermi Liquid phase ($U/t=6$) in Figures $\rm{c})-\rm{d})$. We compare the VQE state-preparation method with the UCCSD ansatz in Eq.~\eqref{eq:uccsd_ansatz} to the VUMPO state-preparation in Eq.~\eqref{eqn:hwp_su4} in the first and second columns, respectively. The mean of the trajectories is shown as a solid curved line, and the standard deviation as a band surrounding the mean. Notably, the standard deviation remains relatively low across the displayed trajectories.
    }
    \label{fig:fermi_hubbard_traj_results}
\end{figure}

Figure~\ref{fig:fermi_hubbard_traj_results} displays the convergence of ground state energy optimizations for both Fermi Liquid (FL) and non-Fermi Liquid (NFL) phases, comparing VQE and VUMPO. The trajectories demonstrate efficient convergence in all cases.

\begin{figure}
    \includegraphics[width=\columnwidth]{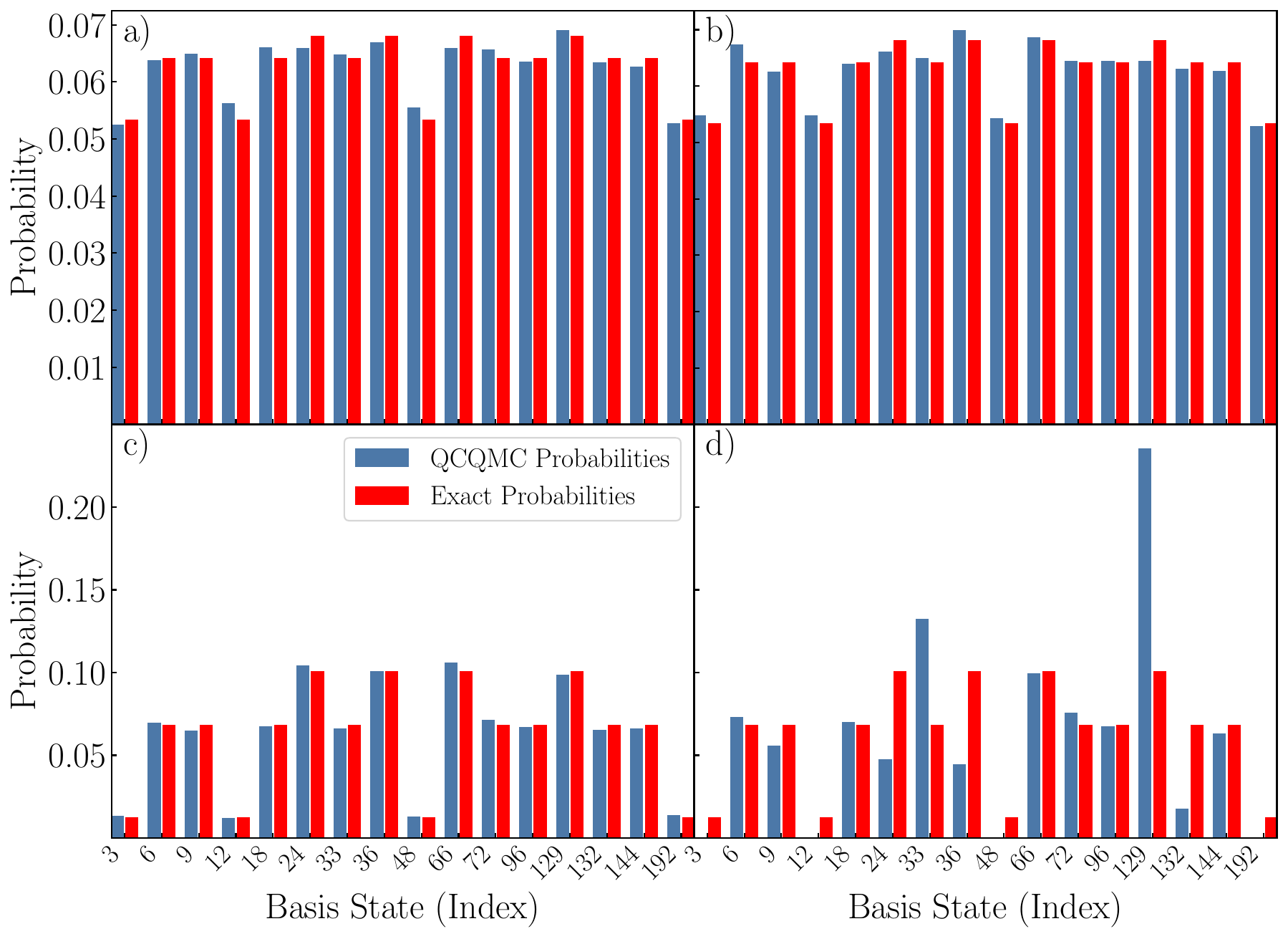}
    \caption{Bitstring basis expansion results for two phases of the spinful \(2\times 2\) Fermi-Hubbard model with QCQMC: the Fermi liquid ($U/t=0.5$) in Figures $\rm{a})-\rm{b})$ and non-Fermi Liquid phase ($U/t=6$) in Figures $\rm{c})-\rm{d})$. Indices correspond to the integer values of the associated bitstrings. We compare the VQE state-preparation method with the UCCSD ansatz in Eq.~\eqref{eq:uccsd_ansatz} to the VUMPO state-preparation in Eq.~\eqref{eqn:hwp_su4} in the first and second columns, respectively. The results from exact diagonalization are also plotted for benchmarking.
    }
    \label{fig:fermi_hubbard_hist_results}
\end{figure}

Figure~\ref{fig:fermi_hubbard_hist_results} presents a detailed state-tomography comparison of the ground state wavefunctions, their bitstring basis expansions, against the results using exact diagonalization. We observe excellent agreement between the VQE-prepared states and the exact solutions. Conversely, VUMPO shows reduced accuracy in the non-Fermi Liquid phase. This discrepancy can be attributed to the increasingly complex and highly entangled nature of the wavefunction in the NFL phase, driven by stronger electron correlations at larger $U/t$ ratios. It is plausible that the underlying tensor-network structure of the VUMPO state preparation approach, especially if configured with a low bond dimension, struggles to efficiently capture these strongly-correlated features.

\subsection{Random Haar Thermal}
To demonstrate the ability of the generalized QCQMC framework to address finite-temperature quantities by means of a Haar pure-state basis, we consider two standard fermionic systems. 

\begin{figure}[t]
    \centering    \includegraphics[width=0.99\columnwidth]{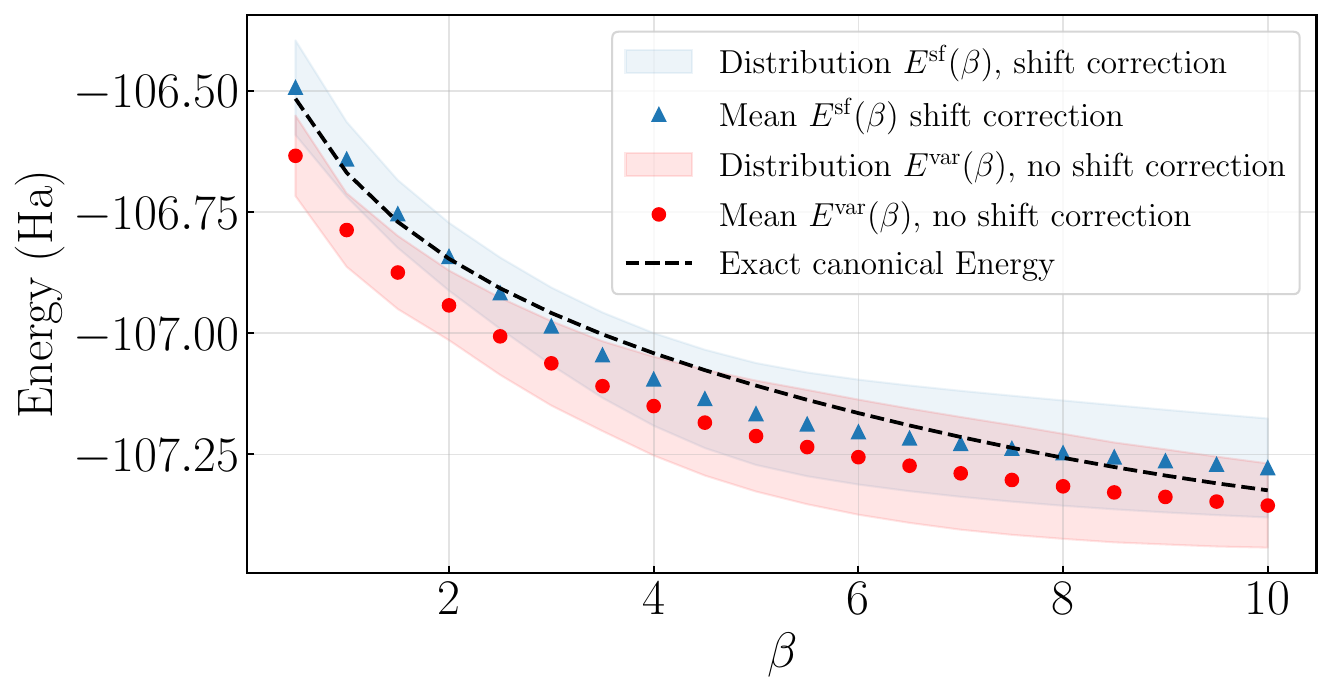}
    \caption{Finite-temperature energy trajectory with QCQMC by means of random Haar $U_\text{Haar}$ state preparation, for a $N_{2}$ molecule, with interatomic distance $0.35$~\AA. In this instance, the system is described by a six qubit Hamiltonian. The exact canonical energy $E(\beta)$ is compared to the QCQMC evaluation using both $E^{(0)}_{\rm{var}}$ and $E^{(0)}_{\rm{sf}}$ estimators. Notice that the color band depict the trajectories for each of the $G=50$ initial random Haar. } 
    \label{fig:haar_n2}
\end{figure}

The models addressed in the previous section, a FH model on a square lattice (of dimensions $2\times3$), without spin degrees of freedom, and a molecular chemistry setting, the $N_{2}$ molecule, with $2$ electrons and $3$ orbitals (hence $6$ qubits in the Jordan-Wigner representation). The approach outlined in this work for addressing finite temperature quantities has some advantages and limitations compared to other approaches to the same problem. For instance, while it is built on pure states and hence in principle does not suffer from the scaling problems of full density matrix approaches, the first-order expansion of the imaginary-time operator, on which QMC is built, limits the achievable accuracy, particularly via the variance of the estimator.  Here, we do not aim to compete or outperform state-of-the-art methods, but rather we want to provide a proof of concept that will serve as a possible benchmark for near-future investigations. 
\begin{figure}[t]
    \centering
    \includegraphics[width=0.95\columnwidth]{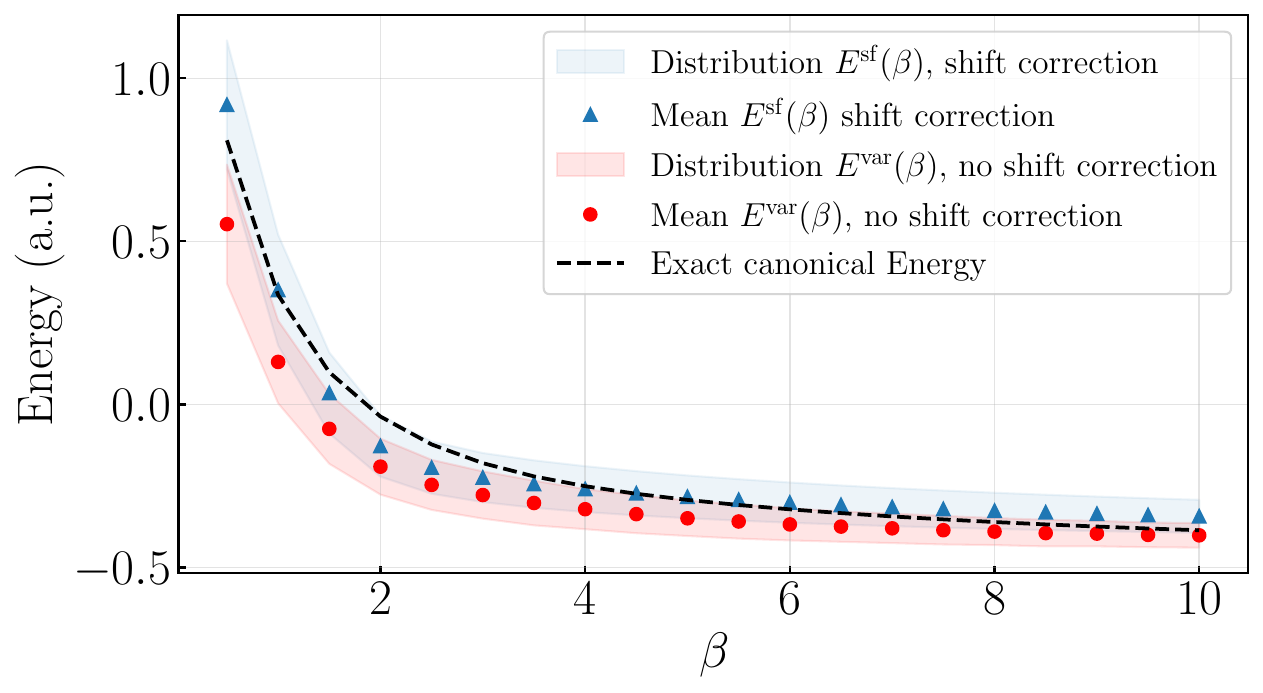}
    \caption{Finite-temperature energy trajectory with QCQMC by the means of random Haar $U_\text{Haar}$ state preparation, for a 2D FH model on a $2\times3$ square lattice, without spin degrees of freedom. Here $U/t=5$ and $\mu=0$. In this instance, the system is described by a eight qubit Hamiltonian. The exact canonical energy $E(\beta)$ is compared to the QCQMC evaluation using both $E^{(0)}_{\rm{var}}$ and $E^{(0)}_{\rm{sf}}$ estimators. Notice that the color band depict the trajectories for each of the $G=50$ initial random Haar. } 
    \label{fig:haar_fh}
\end{figure}
In particular, we estimate via QCQMC, the energy as a function of the inverse-temperature, $\langle E\rangle_{\beta}$. In both systems considered, we sample $G=50$ random Haar unitaries, so that the basis-preparation unitary of the QCQMC pipeline is now a Haar-random unitary, $U_{g}=U_\text{Haar}$. In this way, we realize, averaging over the various Haar realizations, a representation of the maximally mixed state $\rho=\mathbb{I}/\dim(H)$. This is a necessary step for following the thermalization process. Starting from a different initial ensemble would still preserve convergence to the ground state energy (i.e.\ $\beta \rightarrow \infty$), as dictated by the imaginary time evolution, but it would not support, in general, the estimation of finite-temperature quantities, due to the loss of typicality~\cite{Sugiura}.
The collection of $G$ basis states, $\{U_{\text{Haar},g}|b_{i}\rangle=|\widetilde{\psi}_{i,g}\rangle , i=1,\dots,\text{dim}(H)\}^{G}_{g=1}$, will then undergo $G$ independent $\text{QMC}^{(0)}_g$,  realizing at each $\Delta \tau = \Delta\beta/2$ the first order imaginary time step $e^{-\Delta \tau H}|\widetilde{\psi}_{i,g}\rangle \approx(\mathbb{I}-\Delta \tau H)|\widetilde{\psi}_{i,g}\rangle$. Such a set of states, when averaged properly, reproduces the thermal averages from the canonical ensemble, statistically. Notice here that the dynamics of a single pure state would fail to address canonical ensemble quantities, for the imaginary time evolution will only filters, projectively, the information encoded in that specific state. Let us point out here that, for the first-order approximation, the present pipeline will suffer from deviations from the expected value of order $O(\Delta \tau^{2})$. 

If the QMC has a population control strategy in place, i.e. the energy shift $\mathcal{E}^{(m)}(\tau)$ is non zero, then the imaginary time filter might be biased by the energy shift itself, effectively realizing $(\mathbb{I}+\Delta\tau(\mathcal{E}^{(m)}(\Delta\tau)\,\mathbb{I}- H))|\widetilde{\psi}_{i,g}\rangle$. For the thermal averages, taking this quantity into account is crucial: asymptotically in $\tau$, when considering ground or excited state, the effect $\mathcal{E}^{(m)}(\tau)$ can be negligible, as this quantity itself is expected to be proportional to the target energy, i.e., it is a suitable estimator $\overline{f^{(m)}}(H)$. However, this is not true in transient regimes; the shift can significantly impact the statistical distribution of the estimated observables, introducing a bias in the estimator. To mitigate this bias, while still controlling the quantum walker population dynamics, we make use of the estimation strategy $\overline{f^{(m)}}(H)=E^{(0)}_{\rm{sf}}(\tau)$, i.e., the shifted-weighted energy estimator, defined in Eq.~\eqref{eq:en_est_var_sf}.
Results for the fermionic systems are shown in Fig.~\ref{fig:haar_n2} and Fig.~\ref{fig:haar_fh}. In both cases, the energy estimated via $E^{(0)}_{\rm{sf}}$ (blue points) fits the expected behavior better than the standard variational energy $E^{(0)}_{\rm{var}}$ (red points). Notice here that the plots showcase an interesting interplay between two phenomena that minimize the deviation from the theory at low and high $\beta$. At low $\beta$, the linearized version of the imaginary-time evolution implemented via the Monte Carlo is more accurate; at high $\beta$, i.e.\ approaching the ground state, the filtering effect of the quantum walker dynamics is more pronounced.   
An additional source of statistical fluctuation comes from the number of Haar random states needed for reproducing the ensemble: the standard error of the Haar average scales as $1/\sqrt{G}$. Hence, a significant number of Haar unitary circuits needs to be realized to reduce the fluctuations. Notice here that this fact, i.e. the co-occurrence of two sources of error, give rise to rough approximations in some regimes  (see for instance Fig.\ref{fig:haar_n2} at large $\beta$).

\subsection{Cardinality-Constrained MaxCut Optimization}
\label{sec:maxcut}

As mentioned in the previous sections, we now demonstrate that QCQMC can also be applied in combinatorial optimization problems to find optimal bitstring solutions which are encoded in the quantum state obtained at convergence. Table~\ref{tab_graph_descriptions} summarizes the structural properties of the graph instances considered in this work for the optimization problem. For each instance, we show the number of nodes ($n_\mathrm{nodes}$), number of edges, average node degree, and graph density. The graphs are generated according to an Erd\H{o}s--R\'enyi~\citep{erdds1959random} model with fixed edge probability ($p=0.8$), resulting in relatively dense instances that increase in size ($n_\mathrm{nodes} \in \{5, 10, 12, 15\}$).

\begin{table}[h!]
\centering
\label{tab_graph_descriptions}
\begin{tabular}{cccc}
\toprule
$n_\mathrm{nodes}$ & Number of edges & Average degree & Density \\
\midrule
5 & 7 & 2.8 & 0.7 \\
10 & 40 & 8.0 & 0.89 \\
12 & 55 & 9.17 & 0.83 \\
15 & 85 & 11.33 & 0.81 \\
\bottomrule
\end{tabular}
\caption{Summary of the graph instances used for cardinality-constrained MaxCut problems.}
\end{table}

\begin{figure}
\centering
\includegraphics[width=0.48\columnwidth]{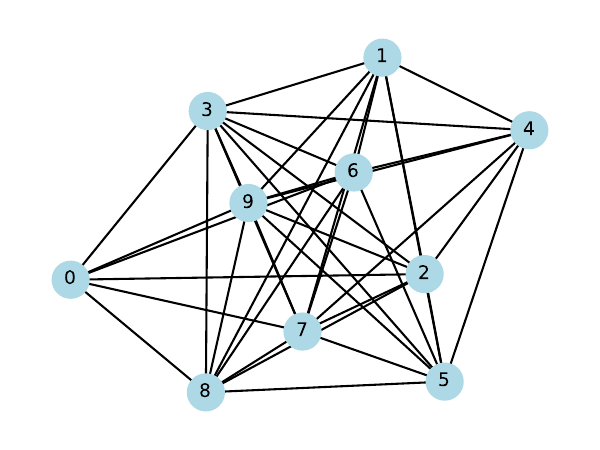}
\includegraphics[width=0.48\columnwidth]{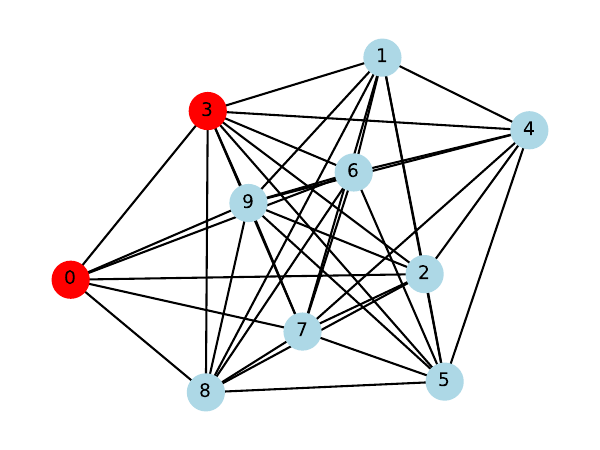}
\caption{Graph visualization and a valid partition for the cardinality-constrained MaxCut optimization problem ($n_{\rm{nodes}} = 10, n_{\rm{ones}}=2$)}
\label{fig:graphs_visualization}
\end{figure}

Figure~\ref{fig:graphs_visualization} illustrates an example graph with $n_{\text{nodes}} = 10$ and unweighted edges. In the cardinality-constrained setting, the size of one of the partitions is fixed to a prescribed value, denoted by $n_{\text{ones}}$. Figure~\ref{fig:graphs_visualization} presents a feasible solution to the cardinality-constrained MaxCut problem with $n_{\text{ones}} = 2$, corresponding to the bitstring $1001000000$, where every labeled node in the graph is represented by a qubit.

Other approaches based on variational circuits, such as QAOA, typically enforce constraints by incorporating penalty terms directly into the Hamiltonian~\citep{liu2025framework}. In the formulation described in Section~\ref{sec_hams_opt}, the optimization must be performed over the full Hamiltonian defined in Eq.~\eqref{eq:hamiltonian}, including both the MaxCut term $H_\mathrm{MaxCut}$ and the constraint contribution $H_\mathrm{c}$ weighted by the penalty parameter $\lambda$. Consequently, the variational procedure must simultaneously optimize the objective while ensuring that violations of the cardinality condition are sufficiently penalized. Other refined approaches include XY mixers~\citep{rieffel2020xy} to satisfy these constraints.

Our QCQMC pipeline employing the L-SPA intrinsically preserves the Hamming weight (representing $n_{\rm{ones}}$ in the graph problem) of basis states throughout the circuit. 
This design enforces the cardinality condition directly at the quantum circuit level, circumventing the need for computationally demanding penalty terms in the Hamiltonian. As a result, substantial computational resources are saved, and the optimization can be performed using only the $H_{\rm{MaxCut}}$ objective, with the constraint inherently satisfied.

\begin{table}
\centering
\label{tab:qaoa_parameters}
\begin{tabular}{cccc}
\toprule
$n_\mathrm{nodes}$ & $p$ & $\lambda$ & Notes \\
\midrule
5  & 2 & 50 & Multiple optimal layers \\
10 & 2 & 10 & Shallow circuit sufficient \\
12 & 2 & 5  & Shallow circuit sufficient \\
15 & 1 & 5  & Single-layer optimal \\
\bottomrule
\end{tabular}
\caption{Recommended QAOA parameters for cardinality-constrained MaxCut. $n_\mathrm{nodes}$ is the number of nodes, $p$ is the QAOA circuit depth, and $\lambda$ is the optimal penalty parameter.}
\end{table}
For comparison, we also evaluate a QAOA-based implementation in which the full Hamiltonian in Eq.~\eqref{eq:hamiltonian} is used, including the penalty term enforcing the cardinality constraint ($\lambda$). Since the performance of this formulation depends on the choice of $\lambda$, we perform preliminary experiments to determine an appropriate value for each problem instance. In particular, we scan over a range of $\lambda$ values and select the configuration yielding the best performance, ensuring a fair comparison between the penalty-based QAOA approach and the symmetry-preserving SPA formulation. Table~\ref{tab:qaoa_parameters} shows the optimal $\lambda$ and $p$ configuration for QAOA approach.

We compare the ideal (noise-free) case with the scenario including readout noise in order to assess the robustness of the QCQMC optimization pipeline and the stability of the resulting solutions. The readout error disturbs the probability that the measured qubit differs from the actual qubit state. It is modeled as a bit-flip channel with error probability $p=0.01$. The corresponding Kraus operators are
\begin{equation}
\begin{split}
    K_0 &= \sqrt{1-p}\,I,\\ 
    K_1 &= \sqrt{p}\,X.
\end{split}
\end{equation}

\begin{figure}
    \includegraphics[width=0.99\columnwidth]{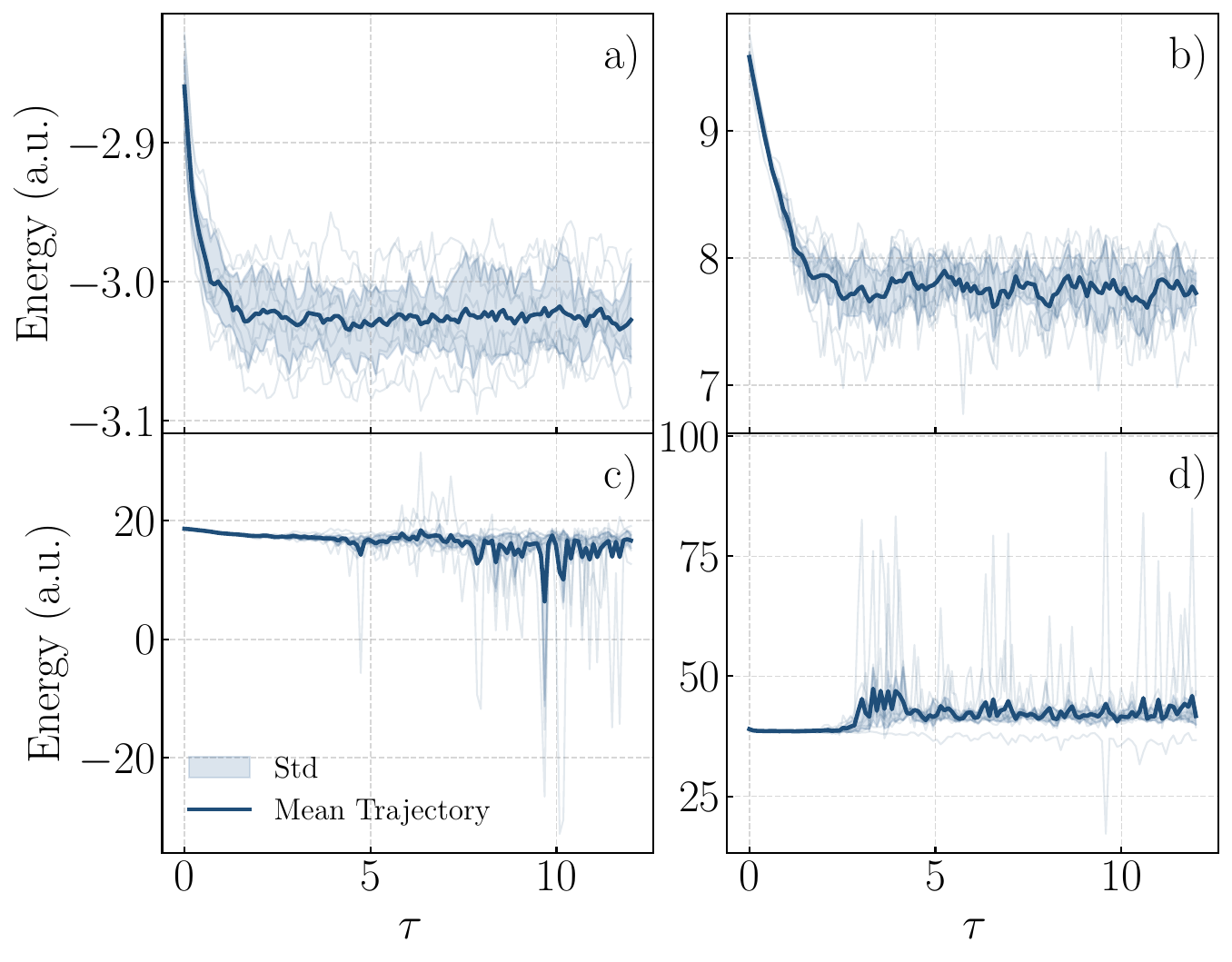}
    \caption{Trajectories of QCQMC with a VQE-prepared state basis with the L-SPA, for the graph-optimization problem with a fixed $n_{\rm{ones}}=2$ constraint as a function of the time propagation (with units of inverse of energy) without readout noise included. Figures $\rm{a})-\rm{d})$ depict the number of nodes $\in \{5,10,12,15\}$ in the graph, respectively. Full details of the problem and hyperparameters are detailed in Sec.~\ref{sec:maxcut}. The optimal solutions are detailed in Table~\ref{tab_results_max_cut}. The mean of the trajectories is shown as a solid curved line and the standard deviation as a band surrounding the mean.}
    \label{fig:opt_conv_no_noise}
\end{figure}

\begin{figure}
    \includegraphics[width=0.99\columnwidth]{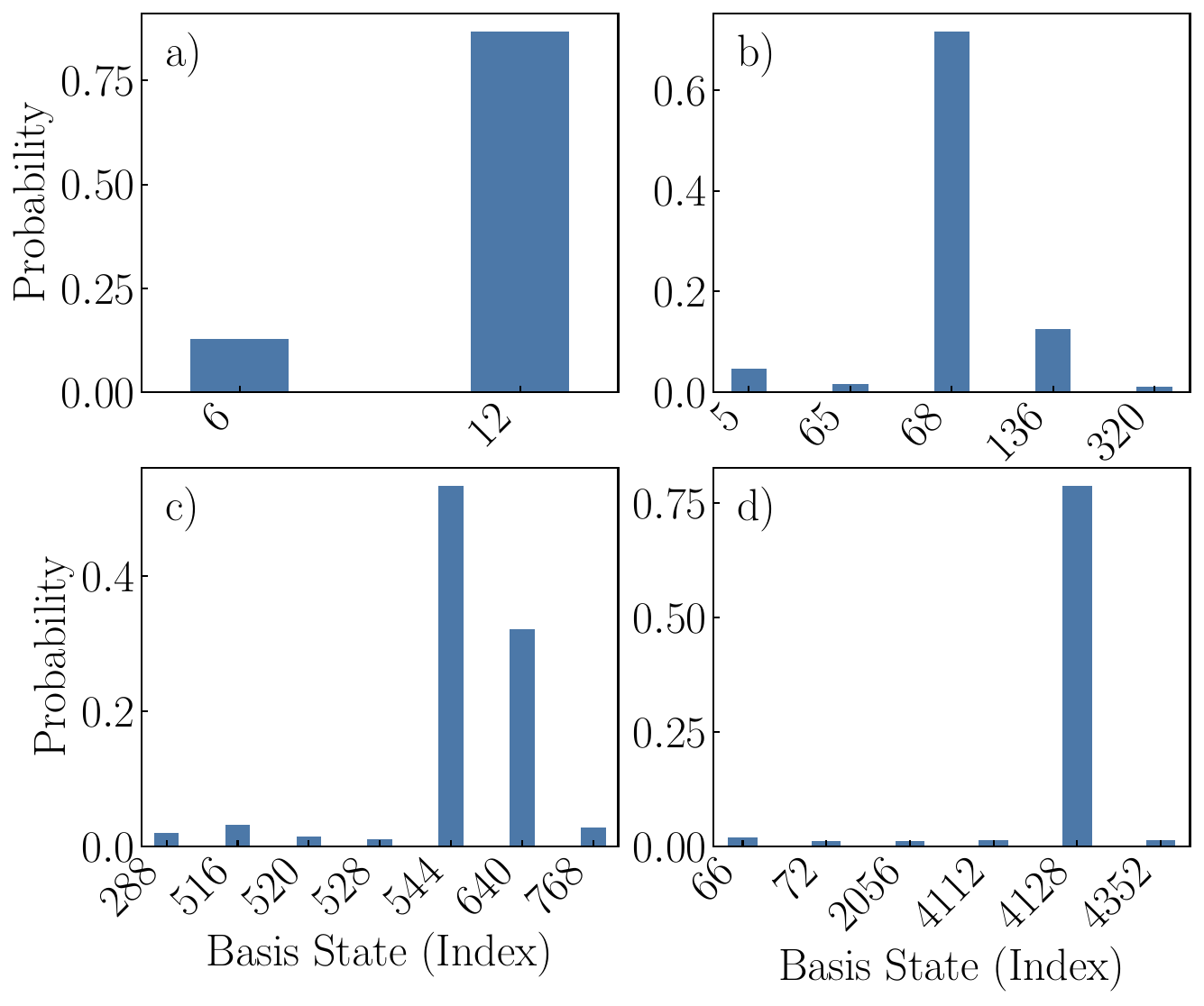}
    \caption{Bitstring basis expansion results for the QCQMC states found without readout noise included. Indices correspond to the integer values of the associated bitstrings. Full details of the problem and hyperparameters are detailed in Sec.~\ref{sec:maxcut}. The optimal solutions for the graph-optimization problem are encoded here as the most likely bitstrings contributing to the quantum state. Figures $\rm{a})-\rm{d})$ depict the number of nodes $\in \{5,10,12,15\}$ in the graph, respectively.}
    \label{fig:hists_no_noise}
\end{figure}

\begin{figure}
    \centering
    \includegraphics[width=0.99\columnwidth]{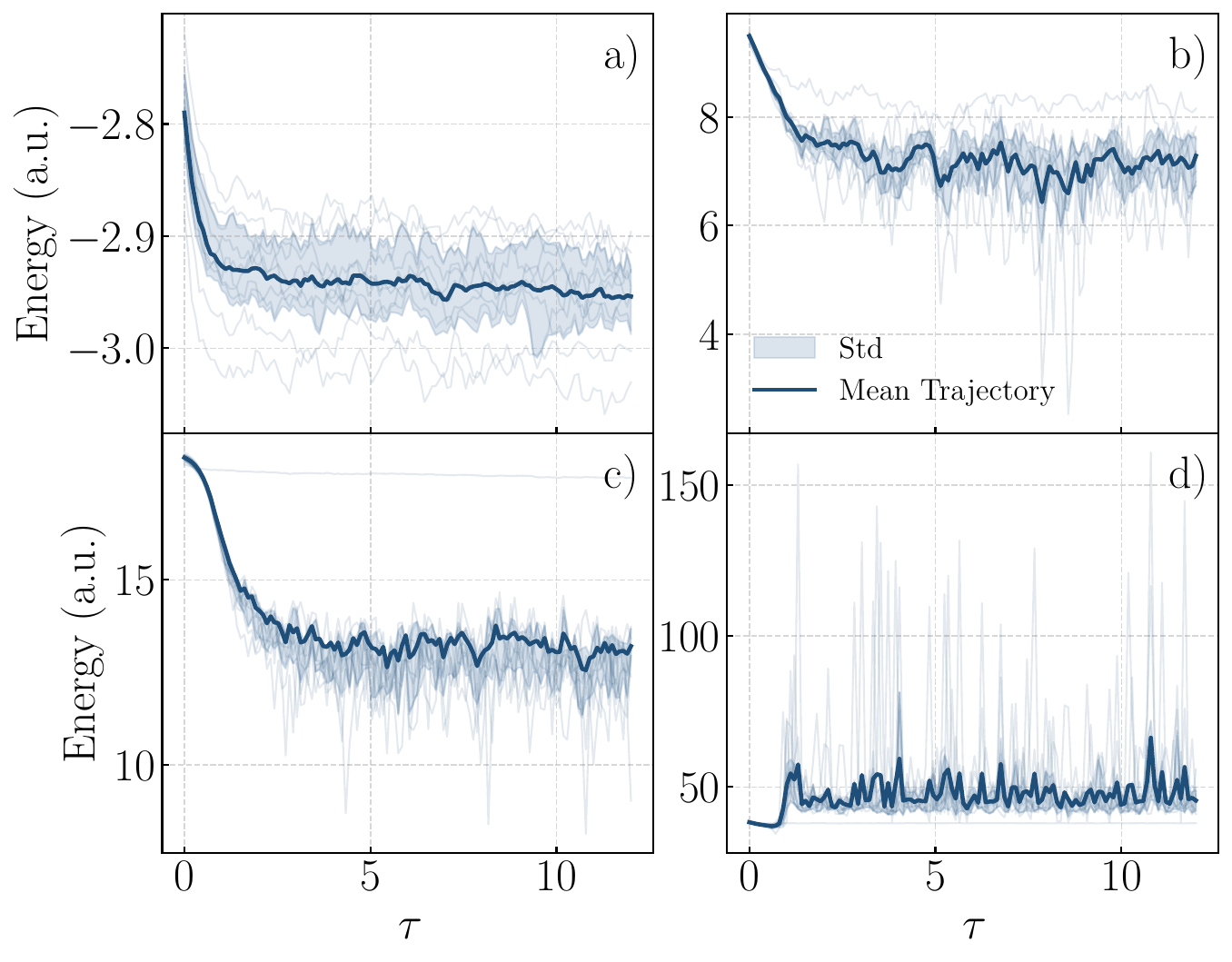}
    \caption{Trajectories of QCQMC with a VQE-prepared state basis with the L-SPA, for the graph-optimization problem with a fixed $n_{\rm{ones}}=2$ constraint as a function of the time propagation (with units of inverse of energy) and with readout noise included. Figures $\rm{a})-\rm{d})$ depict the number of nodes $\in \{5,10,12,15\}$ in the graph, respectively. Full details of the problem and hyperparameters are detailed in Sec.~\ref{sec:maxcut}. The optimal solutions are detailed in Table~\ref{tab_results_max_cut}. The mean of the trajectories is shown as a solid curved line, and the standard deviation as a band surrounding the mean. Even in the presence of noise, the time propagation exhibits remarkable convergence.}
    \label{fig:opt_conv_with_noise}
\end{figure}

\begin{figure}
    \includegraphics[width=0.99\columnwidth]{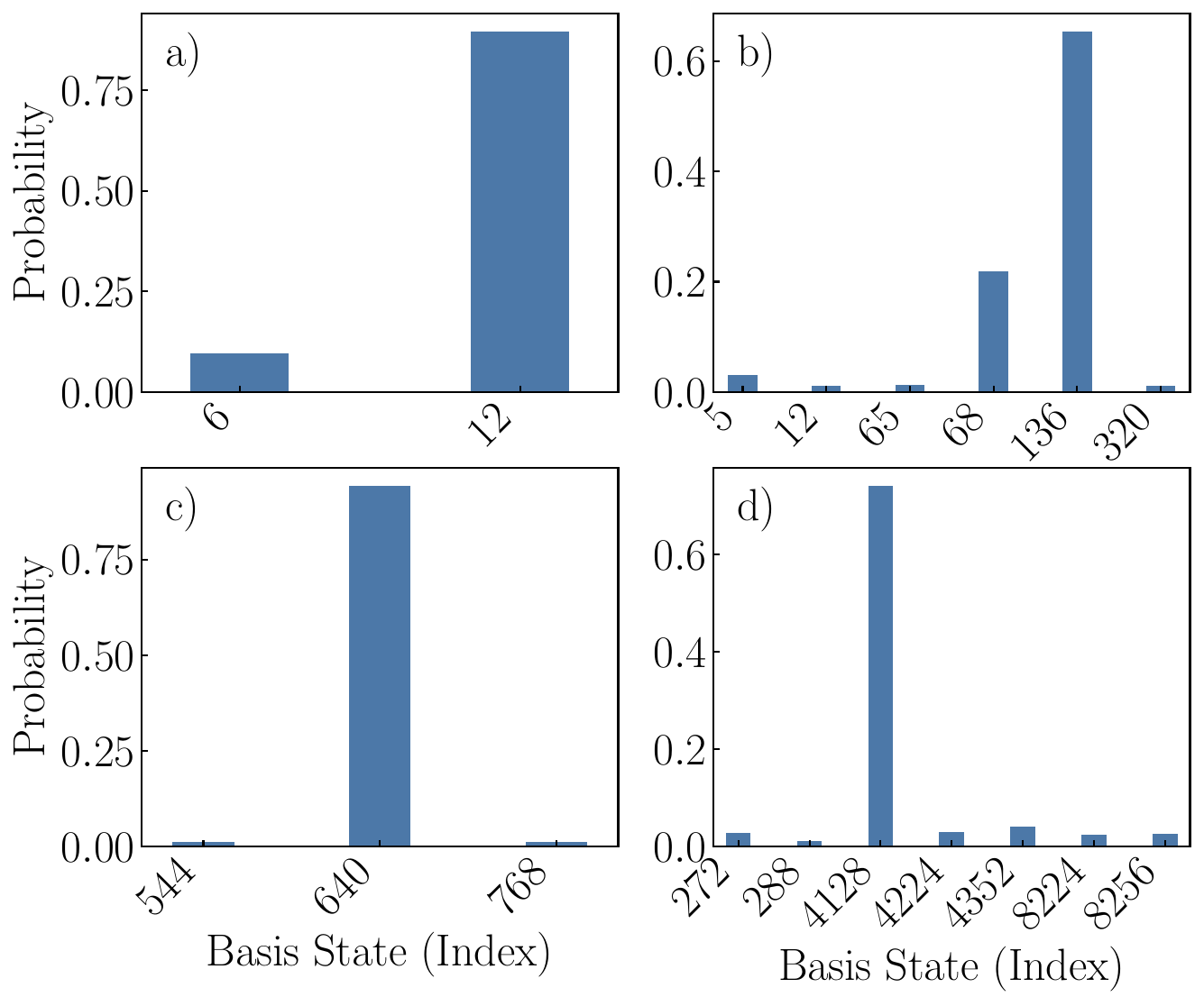}
    \caption{Bitstring basis expansion results for the QCQMC states found with readout noise included. Indices correspond to the integer values of the associated bitstrings. Full details of the problem and hyperparameters are detailed in Sec.~\ref{sec:maxcut}. The optimal solutions for the graph-optimization problem are encoded here as the most likely bitstrings contributing to the quantum state. Figures $\rm{a})-\rm{d})$ depict the number of nodes $\in \{5,10,12,15\}$ in the graph, respectively. Despite the noise, the optimization still concentrates probability on desired solutions.}
    \label{fig:hists_with_noise}
\label{fig_results_opt_noisy}
\end{figure}

Table~\ref{tab_results_max_cut} summarizes the results obtained for the cardinality-constrained MaxCut problem with $n_{\rm{ones}}=2$ for different graph sizes. The performance of the QCQMC approach is compared against the optimal solutions obtained via brute-force search and against the best results obtained from the penalty-enhanced QAOA. The table reports the best cut values found in each case, both in the ideal scenario and in the presence of the previously-described readout noise. Figures~\ref{fig:opt_conv_no_noise} and~\ref{fig:hists_no_noise} show the convergence plot and corresponding histogram in a noise-free setting that yield the QCQMC results summarized in Table~\ref{tab_results_max_cut}. The QCQMC pipeline was run for $2,000$ shots with VQE-prepared basis with a unit-layer L-SPA ansatz. We use an average over ten trajectories with a time step of $0.1$ (in units of inverse of energy) for 120 iterations. The number of initial walkers and the threshold are $1,000$ and $200$, respectively.
The corresponding Figs.~\ref{fig:opt_conv_with_noise} and~\ref{fig:hists_with_noise} show the same results, including readout noise. 


\begin{table}[h!]
\centering
\label{tab_results_max_cut}
\begin{tabular}{ccccc}
\toprule
$n_\mathrm{nodes}$ & Brute Force & QAOA (best) & \multicolumn{2}{c}{QCQMC} \\
\cmidrule(lr){2-2}
\cmidrule(lr){4-5}
 & Optimal Cut & & Noise-free & Noisy \\
\midrule
5  & 5  & 5  & 5  & 5 \\
10 & 16 & 12 & 16 & 16\\
12 & 19 & 18 & 19 & 19\\
15 & 24 & 24 & 24 & 24\\
\bottomrule
\end{tabular}
\caption{Comparison of solution quality for cardinality-constrained MaxCut ($k=n_{\rm{ones}}=2$).}
\end{table}

As shown, QCQMC consistently recovers the optimal cut values across all tested instances. Moreover, the inclusion of readout noise does not degrade the solution quality in these experiments, indicating that the method exhibits a strong degree of robustness with respect to measurement errors. Its noise resilience and the fact of not considering the constrained part of the Hamiltonian suggest that this approach may be a good fit for larger-scale problem instances in which the constraint can be encoded into the Hamming weight of the ansatz.

\subsection{Nuclear shell model}
\label{sec:nsm}

\begin{figure}
    \centering
    \includegraphics[width=0.95\columnwidth]{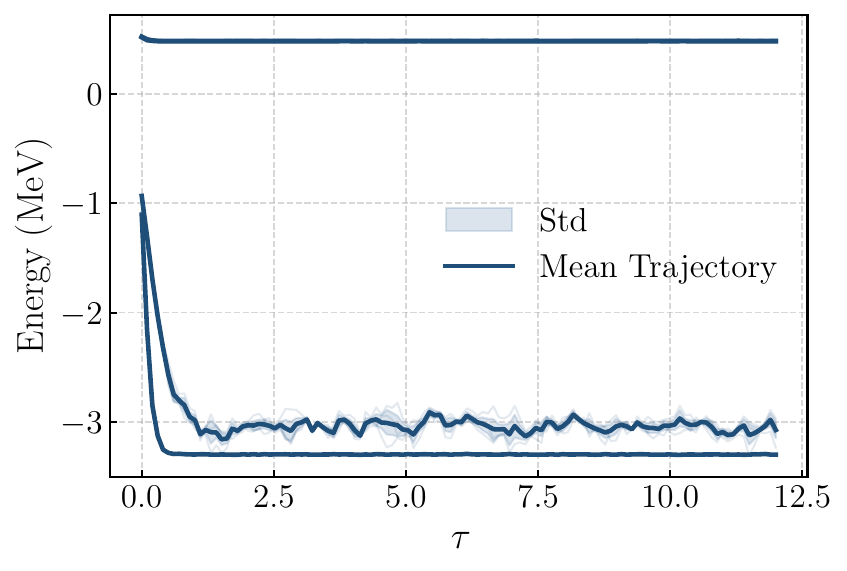}
    \caption{Trajectories of QCQMC with a VUMPO-prepared state basis for the ground- and low-lying excited states for the nuclear shell model Hamiltonian in the \emph{p-}shell. The mean of the trajectories is shown as a solid curved line, and the standard deviation as a band surrounding the mean. Notably, the standard deviation remains relatively low across the displayed trajectories.}
    \label{fig:nsm_vumpo}
\end{figure}

\begin{figure}
    \centering
    \includegraphics[width=0.9\columnwidth]{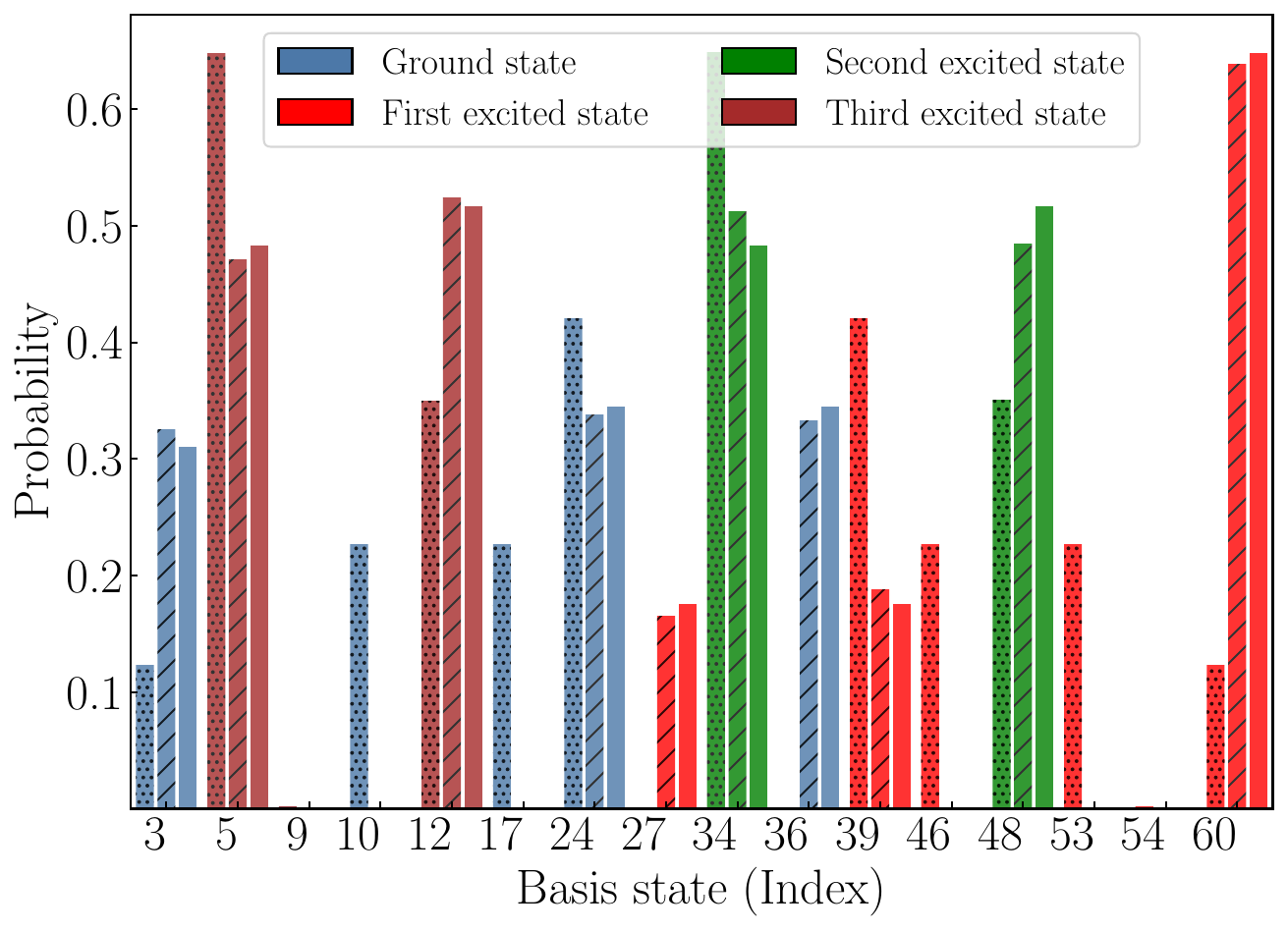}
    \caption{Bitstring basis expansion for the ground- and low-lying excited states of the nuclear shell model Hamiltonian in the \emph{p-}shell for the VUMPO state-preparation method (dotted bars) and QCQMC corrections (solid bars). Indices correspond to the integer values of the associated bitstrings. Results from exact diagonalization (dashed bars) are also plotted for benchmarking. The results span over multiple nuclei, with further details summarized in Table~\ref{tab:nsm_amplitude_results}.}
    \label{fig:nsm_vumpo_hist}
\end{figure}

In this work, we investigate the nuclear shell model within two distinct valence spaces: the $p$-shell and the $sd$-shell, as illustrated in Fig.~\ref{fig:nsm_configuration_space}. We employ the Jordan-Wigner transformation to map fermionic nuclear states onto qubit states.  
The QCQMC pipeline offers the flexibility to simulate multiple nuclei within the chosen valence space in a single computational run by disregarding the Hamming-weight symmetry. Thus, excited states may correspond to ground states of different nuclei according to the Hamming-weight of the corresponding basis state (Slater determinant) expansion. We utilize effective Hamiltonians derived from the Cohen-Kurath~\cite{Cohen_Kurath} and the mass-independent USDB~\cite{newusd} interactions for the $p$- and $sd$-shells, respectively. Our only constraint is to consider either protons or neutrons, but not both simultaneously, in a given simulation. This decision leads to six qubits for the $p$-shell and twelve for the $sd$-shell, a valid simplification since the employed effective interactions used possess isospin symmetry, meaning they are invariant with respect to the exchange of protons and neutrons.

Figure~\ref{fig:nsm_vumpo} shows the time-propagated trajectories for four low-lying eigenstates of the $p$-shell nuclear Hamiltonian using a VUMPO state-preparation approach. We use an average over three trajectories with a time step of $0.1$ for $120$ iterations. The number of initial walkers and threshold are $2,000$ and $500$, respectively, to ensure convergence. The corresponding basis state expansion and comparison with the results obtained from exact diagonalization methods are plotted in Fig.~\ref{fig:nsm_vumpo_hist} and summarized in Table~\ref{tab:nsm_amplitude_results}, where we can see that it spans over multiple nuclei in the valence space. The strong agreement observed demonstrates that QCQMC can reliably extract basis state amplitudes, provided the quantum state exhibits sufficient sparsity. 

\begin{figure}[t]
    \centering
    \includegraphics[width=0.95\columnwidth]{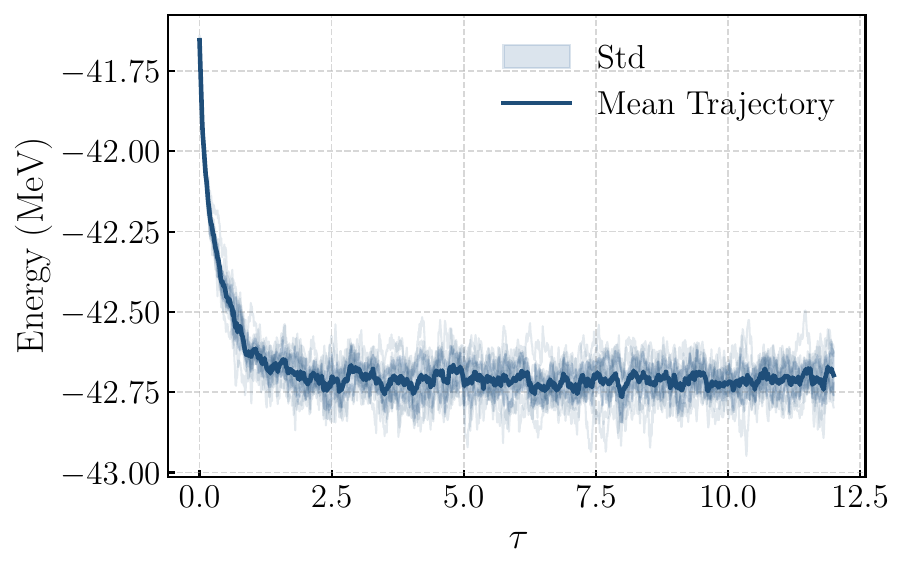}
    \caption{Energy convergence of the ground-state for the nuclear shell model Hamiltonian in the \emph{sd-}shell. The mean of the trajectories is shown as a solid curved line and the standard deviation as a band surrounding the mean.}
    \label{fig:nsm_sdshell}
\end{figure}

\begin{table} 
    \centering
    \begin{tabular}{@{} c c S[table-format=1.4] S[table-format=1.4] @{}} 
        \toprule
        \textbf{Target State} & \textbf{Basis State} & \multicolumn{2}{c}{\textbf{Probabilities}} \\
        \midrule
        Ground state & 011000 & 0.3379 & 0.3447 \\
          & 100100 & 0.3331 & 0.3447 \\
          & 000011 & 0.3251 & 0.3107 \\
        \midrule 
        Excited state 1 & 111100 & 0.6389 & 0.6483 \\
          & 100111 & 0.1881 & 0.1758 \\
          & 011011 & 0.1651 & 0.1758 \\
        \midrule
        Excited state 2 & 100010 & 0.5124 & 0.4832 \\
          & 110000 & 0.4845 & 0.5168 \\
        \midrule
        Excited state 3 & 001100 & 0.5237 & 0.5168 \\
          & 000101 & 0.4710 & 0.4832 \\
        \bottomrule
    \end{tabular}
    \caption{Comparison between the Monte Carlo simulated basis state probabilities (left) with the ones obtained from exact diagonalization (right). The found ground state and excited states 2 and 3 correspond to nuclear states of $^6$Be (or $^6$He, given the proton-neutron symmetry) and excited state 1 corresponds to the nuclear ground state of $^8$C (or $^8$He). }
    \label{tab:nsm_amplitude_results}
\end{table}
Figure~\ref{fig:nsm_sdshell} shows the time-propagated trajectories for the ground state of the $sd$-shell nuclear Hamiltonian using a VQE state-preparation approach with the UCCSD ansatz in Eq.~\eqref{eq:uccsd_ansatz}, where a single Trotter step and second Trotter order were used. We use an average over five trajectories with a time step of $0.01$ for $1200$ iterations. The number of initial walkers and the threshold are $1,000$ and $100$, respectively. The corresponding basis state expansion and comparison with the results obtained from exact diagonalization methods are summarized in Table~\ref{tab:sd_amplitude_results}. Similarly as what was observed formerly for the $p-$shell, there is a strong agreement with the results obtained from exact diagonalization in this case, even for a very different set of hyperparameters in the QCQMC pipeline, demonstrating the robustness of the method. 

\begin{table}[htb!] 
    \centering
    \begin{tabular}{@{} c c S[table-format=1.4] S[table-format=1.4] @{}} 
        \toprule
        \textbf{Target State} & \textbf{Basis State} & \multicolumn{2}{c}{\textbf{Probabilities}} \\
        \midrule 
        Ground state &  011110111111 & 0.0138 & 0.0152 \\
         & 101101111111  & 0.0210 & 0.0152 \\
         & 110011111111  & 0.0171 & 0.0152 \\
        & 111111001111 & 0.0180 & 0.0197 \\
        & 111111110110 &  0.4721 & 0.4673 \\
        & 111111111001 & 0.4564 & 0.4673 \\
        \bottomrule
    \end{tabular}
    \caption{Comparison between the Monte Carlo simulated basis state probabilities (left) with the ones obtained from exact diagonalization (right). The ground state corresponds to the ground state of $^{26}$Ar (or $^{26}$O). }
    \label{tab:sd_amplitude_results}
\end{table}

\section{Discussion and conclusions}\label{sec:conclusion}

In this work, we have introduced and systematically bench-marked a unified Quantum Computing Quantum Monte Carlo (QCQMC) framework that significantly expands its applicability beyond typical ground state energy calculations. By demonstrating the versatility of QCQMC across diverse computational domains—including molecular chemistry, the Fermi-Hubbard model, the nuclear shell model, and cardinality-constrained MaxCut optimization—we have shown its potential to address a broader range of complex problems in quantum simulation.

A key contribution of this framework is the methodical variation of the unitary operator ($U_g$) used for basis state generation. We explored several quantum state preparation strategies, each tailored to specific problem characteristics and computational objectives. For estimating ground and excited states, variational quantum algorithms such as VQE and Variational Fast Forwarding (VFF) were employed. VQE demonstrated good performance in certain scenarios. When compared for the problem considered, VQE and VFF performed equivalently at targeting higher-energy states. Though we believe VFF would outperform VQE for excited states for larger scale problems, computational resource requirements prohibited this, hence the consideration of VUMPO. The Variational Unitary Matrix Product Operator (VUMPO) approach consistently outperformed both VQE and VFF in accurately estimating eigenstate energies. This highlights VUMPO's high fidelity and its ability to effectively handle complex wavefunctions, often making the corrections provided by QCQMC negligible.

For finite-temperature averages, the use of Haar random unitaries demonstrated the framework's ability to estimate thermal quantities from pure-state dynamics. This approach, while distinct from full density matrix methods, offers advantages in terms of scaling. However, it also presents challenges related to the accuracy of the estimator variance and the number of Haar random states required to reduce statistical fluctuations. The incorporation of a shifted-weighted energy estimator ($E_{\rm{sf}}(\beta)$) proved crucial in mitigating bias and improving accuracy, particularly in transient regimes and at higher temperatures.

The application of QCQMC to cardinality-constrained MaxCut optimization problems further showcased its flexibility. By intrinsically preserving Hamming weight symmetry through the layered symmetry-preserving ansatz, our approach effectively enforced constraints at the quantum circuit level, avoiding computationally demanding penalty terms. This not only saved significant computational resources but also demonstrated robustness against measurement errors, indicating suitability for larger-scale problem instances.

A significant challenge in QCQMC arises from the quantum-circuit evaluation of Hamiltonian matrix elements as in Figs.~\ref{fig:modHij_U} and~\ref{fig:P_overlap}. While quantum computers offer a natural environment for sampling these terms, the number of distinct quantum circuits required to estimate all matrix elements, necessary for the population dynamics in QCQMC, can grow exponentially intractable if the Hamiltonian or the basis expansion of quantum ansatz are not sufficiently sparse. This exponential scaling stems from the need to measure numerous individual terms in across all basis states in the expansion of the ansatz, quickly overwhelming even quantum resources. However, this critical bottleneck could be substantially alleviated by a strategic approach: first, leveraging the state preparation capabilities to generate an ansatz\"{e} in a significantly more compact and relevant basis; and second, concurrently employing a judiciously truncated Hamiltonian comprised only of the most physically-significant terms. By ensuring that the Hamiltonian~\eqref{eq:lcu_ham} is polynomially bounded in terms of the number of Pauli strings it comprises, we inherently limit the connectivity within the Hilbert space; that is, any given basis state can only directly interact with a polynomially-sized number of other basis states under the action of this Hamiltonian. When this is coupled with working within a compact, polynomially-sized active basis (e.g., derived from effective state preparation as in Eq.~\eqref{eq:walker_transformation}), the total number of distinct Hamiltonian matrix elements remains polynomially bounded. Consequently, the overall computational complexity, specifically the number of quantum circuits needed for Hamiltonian matrix elements evaluation within the QCQMC framework, is kept under polynomial control, making the simulation tractable.

Future research directions include further optimizing the state preparation protocols for specific problem classes, exploring advanced error mitigation techniques for enhanced accuracy and scalability on noisy intermediate-scale quantum (NISQ) devices, and investigating the integration of quantum machine learning techniques for more adaptive and efficient basis state generation. Ultimately, this unified QCQMC framework represents a promising path towards unlocking the full potential of quantum computers for the simulation and characterization of complex quantum and classical systems.

\begin{acknowledgments}
The authors thank all the members of the Fujitsu Research of Europe team for their helpful comments.
\end{acknowledgments}

\bibliography{references_qmc}

\appendix

\section{Full Configuration Interaction Quantum Monte Carlo}
\label{sec:fciqmc}
In its classical computational form, i.e., without quantum circuits involved, Full Configuration Interaction Quantum Monte Carlo (FCIQMC)~\cite{booth2009fermion,cleland2012taming} is defined as a stochastic diffusion technique designed for inferring the ground state energy of a quantum system. It works by projecting the initial state dynamically onto the ground state through the probabilistic application of the first order expansion of the imaginary-time evolution operator.  Let us consider a time-dependent wavefunction, expanded onto a given basis:
\begin{equation}
|\Phi(\tau)\rangle = \sum_{k} c_k(\tau) |b_k\rangle,
\end{equation}
where  $|b_k\rangle$ could denote Slater determinants in Fock space for the description of fermionic systems~\cite{giuliani2025precise}, and $c_k$ are the coefficients of the linear combination over the basis states, which carry the time dependency. The dynamics of this wavefunction will be modeled via the imaginary-time Schr\"odinger equation~\cite{GOLDBERG1967}:
\begin{equation}
(H-\mathcal{E}\,\mathbb{I}) |\Phi\rangle = - \frac{d}{d \tau} |\Phi\rangle,
\end{equation}
where $\mathcal{E}\,\mathbb{I}$ is an effective shift of the energy that guarantees the normalization, lost by the lack of unitarity \cite{Vigor2015}. The real coefficient $\mathcal{E}$, can be dynamically updated, as we shall see in the following, thus serving as a form of bias control in the stochastic evolution. 
It is worth noting here, for the relevance of the matter in the case of thermal averages, that the equation above defines a non-unitary thermalization process. This is better described by the operator $U=e^{-\tau H}$: this operator prepares a thermal state from the Gibbs ensemble with inverse temperature $\tau$.  Hence, at large imaginary time, the system is forced to converge to the ground state.  In this sense, the FCIQMC can  be interpreted as a first order expansion of the imaginary time evolution operator, effectively realized via random sampling in a imaginary time step $\Delta \tau$. 

Expanding the equation above for the coefficients, we obtain,
\begin{equation}
\label{eq:coeff_dyn}
\frac{d c_r}{d \tau} = -(H_{r} - \mathcal{E}(\tau)) c_r + \sum_{r \neq s} H_{rs} c_s,
\end{equation}
where $H_{rs} = \langle b_r | H | b_s \rangle$ are the Hamiltonian transition matrix elements on the basis states and $ \mathcal{E}(\tau)$ is  again the energy shift, which is assumed to change over time to stabilize the dynamics of the coefficients. The second term on the right-hand side of the equation connects the $r-$th component of the wavefunction expansion to the $s-$th components, while the first term models the survival of the $r-$th component. Discretizing the equation above in $\Delta\tau$, and considering an initial distribution of $N_r$ walkers in the state $|b_r\rangle$, the equation of motion for the coefficients directly induces a probabilistic interpretation: the probability of \textit{spawning}  a new walker $s$ starting from each $r$  is proportional to the off-diagonal elements of the Hamiltonian, while the probability of \textit{killing or cloning} (i.e. increasing its relative weight) depends on the shifted diagonal energies. In particular, if $H_{r}-\mathcal{E}(\tau)\ge0$, the walker $r$ will be cloned with a certain probability; on the contrary it will be suppressed. In a formal way these facts are represented by (up to a normalization factor) the probabilities:
\begin{equation}\label{eq:probs_walkers}
\begin{split}
p^{sp}_{r \to s} &= \Delta\tau H_{rs}, \\
p^{d/k}_{r} &= \Delta\tau (H_{r} - \mathcal{E}(\Delta\tau)),
\end{split}
\end{equation}
where $\Delta\tau$ is the time step used in the discretizations of the basis coefficients dynamics as in \eqref{eq:coeff_dyn}. Thus, for each of the initial $N_r(0)$ walkers  $|b_r\rangle$, new walkers are spawned and/or the starting ones are killed or cloned. The relative number of each walker at time step $\Delta\tau$ represents the real part of the updated coefficients of the linear combination defining the overall quantum state. What is left is the definition of the sign of each spawned walker, i.e., its relative complex phase. Following the standard prescription of the algorithm, the sign of the spawned walker $s$ is dictated by the sign of $H_{rs}$: if the transition element is positive, the child walker inherits the same sign as the parent, if negative the child walker sign is given by $-\frac{H_{rs}}{|H_{rs}|}$.  Given the statistical nature of the estimation, the same walker can pick up different signs during the spawning process. Hence, for guaranteeing the consistency (a walker can have only a single phase per time step), an \textit{annihilation} step is performed, which consists of canceling identical walkers with opposite signs. Combining these ingredients, the wavefunction at time step $\Delta \tau$ is given by:
\begin{equation}
    |\Phi(\Delta\tau)\rangle=\frac{1}{\sqrt{\sum_{r=0}^{\text{dim}(H)}N_{r}^{2}(\Delta \tau)}}\sum_{r}\text{sign}[N_{r}]N_{r}(\Delta \tau)|b_{r}\rangle.
\end{equation}
The estimation of the energy can be performed using various strategies, depending on the complexity of the systems, the target, and the presence of possible sources of bias. An unbiased estimator for the ground state energy, used thoroughly when dealing with this target in the present work, is the projective energy estimator. Let us consider $N_r(0)$ copies of the initial walker $|b_r\rangle$ at $\tau=0$, with energy $\mathrm{E}_{r}$, which its likely to have a good overlap with the target ground state. Then the projective energy estimator takes the form of a correction over the initial energy, namely:
\begin{equation} 
E^{0}_{\rm{pr}}(\tau) =\frac{\langle b_{r }|H|\Phi(\tau)\rangle}{\langle b_{r} |\Phi(\tau)\rangle} = E_{r} + \sum_{i \neq r} \frac{\textrm{sign}(b_i)N_i(\tau)}{N_{r}(0)} H_{i, r}, \label{eq:en_est} \end{equation} 
with $\lim_{\tau \to \infty}E^{0}_{\rm{pr}}(\tau)=E_{\rm{GS}}$. 

How well the initial walker state overlaps with the true ground state, $\langle \Psi_{GS}|b_{r}\rangle$, is extremely important for this algorithm. Just like in standard imaginary time evolution, this overlap determines how quickly the method reaches a solution and how many walkers are needed.  Or, more specifically, the filtering rate of the imaginary time evolution is determined by the initial overlap. This strong dependence highlights QCQMC's advantage in inherently managing the quality of initial states, a problem often discussed in quantum state preparation literature~\cite{fomichev2024initial,zhang2021low}.
The energy shift factor ($\mathcal{E}$) influences, as it enters the death-cloning term, how walkers proliferate or are being removed, thus allowing to control their population. If too many walkers appear, we can adjust $\mathcal{E}$ to stop their numbers from growing uncontrollably, preventing computational overload. We acknowledge that $\mathcal{E}$ is mostly an educated guess, but we used a standard value based on the energies obtained in the state-preparation processes.

The same pipeline described above can be adapted to estimate excited states~\cite{blunt2015excited}. Let us suppose we are aiming at the  $m=0...M-1$ energy levels of the spectrum of a certain system.  The usage of FCIQMC for this problem requires running $M$ algorithms in series, each targeting an individual excited state. 
The simulations entail an orthogonalization step that is performed right after the annihilation step in the single state FCIQMC. This orthonormalization is performed with respect to the instantaneously estimated lower energy states: for instance, the wavefunction for the first excited state is estimated orthorgonalizing respect to the walker distribution for the ground state, the second excited state with respect to the first and ground distribution, and so on. In general and more formally,  for obtaining a walker distribution describing the $m^{\rm{th}}$ excited state $|\Phi^m(\tau)\rangle$, with $m>0$ the ansatz is orthogonalized with respect to the wavefunctions for the lower energy states via the projectors:
\begin{align}
\label{eq:app_orton}
    |\Phi^{(m)}(\tau)\rangle \longrightarrow \left(\mathbb{I} - \sum_{l<m}\frac{|\Phi^{(l)}(\tau)\rangle \langle \Phi^{(l)}(\tau)|}{\langle \Phi^{(l)}(\tau)|\Phi^{(l)}(\tau)\rangle} \right)|\Phi^{(m)}(\tau)\rangle,
\end{align}
and then $|\Phi^{(m)}(\tau)\rangle$ undergoes the usual first order imaginary time step of the dynamics, i.e.  $|\Phi^{(m)}(\tau+\Delta\tau)\rangle=(\mathbb{I}-\Delta\tau(H-\mathbb{I}\mathcal{E}_{s}))|\Phi^{(m)}(\tau)\rangle$.
Consequently, the different FCIQMCs calculations are minimizing the energy of all approximated states whilst enforcing orthogonalization, leading to the preparation of the $l$ lowest eigenstates.

It is important to note that the energy estimator~\eqref{eq:en_est} generally performs poorly when considered for excited states. This is because the accuracy of this estimator scales with the proportion of walkers that remain in the initial state; this proportion is often lower for excited states in QCQMC due to the effects of the projection step. The higher in energy is the target excited state in fact, the more likely is the estimated initial overlap to be small, hindering the QMC diffusion process. Various other estimators can be used, such as that presented in~\cite{petruzielo2012semistochastic} or directly evaluating $\langle \Phi^{(m)}| H |  \Phi^{(m)}\rangle$.

\section{Modified Hadamard test quantum circuits}\label{sec:mod_had_circ}
\begin{figure}[htbp]
    \centering
    \includegraphics[width=0.8\columnwidth]{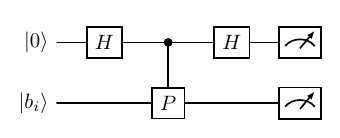}
    \caption{Modified Hadamard test to compute the absolute value of all the transition matrix elements $|H_{ij}| = |\langle b_j|H|b_i \rangle|$ for a fixed input state $|b_i\rangle$.}
    \label{fig:modHij_noU}

\end{figure}

The circuit as depicted in Fig.~\ref{fig:modHij_noU} allows us to discern the relevant transitional states post-processing the measurement results. 
Using ancilla $a$ and state $s$ registers, its action is as follows:
\begin{itemize}
\item The initial state in the circuit is
\begin{equation}
    |\Psi_0\rangle = |0\rangle_a |b_i\rangle_s.
\end{equation}
\item Apply a Hadamard to the ancilla,
\begin{equation}
    |\Psi_1\rangle
    =
    \frac{1}{\sqrt{2}}
    \left( |0\rangle_a + |1\rangle_a \right)|b_i\rangle_s.
\end{equation}

\item Apply a controlled-$P$ gate,
\begin{equation}
    |\Psi_2\rangle
    =
    \frac{1}{\sqrt{2}}
    \left(
        |0\rangle_a\,|b_i\rangle_s
        +
        |1\rangle_a\,P|b_i\rangle_s
    \right),
\end{equation}
where, expanding the action of $P$,
\begin{equation}
    P|b_i\rangle = \sum_{j} P_{ji}\,|b_j\rangle,
\end{equation}
we obtain
\begin{equation}
    |\Psi_2\rangle
    =
    \frac{1}{\sqrt{2}}
    \sum_{j}
    \left(
        |0\rangle_a\,\delta_{j i}
        +
        |1\rangle_a\,P_{ji}
    \right)
    |b_j\rangle_s,
\end{equation}
where we have used that $\sum_j \delta_{ij} |b_j\rangle = |b_i\rangle$

\item Apply a second Hadamard gate to the ancilla,
\begin{equation}
    |\Psi_3\rangle
    =
    \frac{1}{2}
    \sum_{j}
    \Big[
        \big( \delta_{j i} + P_{ji} \big) |0\rangle_a
        +
        \big( \delta_{j i} - P_{ji} \big)|1\rangle_a
    \Big]
    |b_j\rangle_s .
\end{equation}

\item We now measure all qubits in the computational basis. Let $p(a,j)$ denote the joint probability of measuring
ancilla $a\in\{0,1\}$ and state $j$. From the above state,
\begin{equation}
\begin{split}
    p(0,j) & =
    \frac{1}{4}\,\big|\delta_{j i} + P_{ji}\big|^2,
    \\
    p(1,j) & = \frac{1}{4}\,\big|\delta_{j i} - P_{ji}\big|^2.
\end{split}
\end{equation}
Thus, there are two cases to consider. In the case $i \neq j$, we obtain the following:

\begin{equation}
\begin{split}
    p(0,j) &= p(1,j) = \frac{1}{4} \big| P_{ji}\big|^2 \\ |P_{ji}| &= 2 \sqrt{p(0,j)} = 2 \sqrt{p(1,j)}.
\end{split}
\end{equation}

In the case $i = j$, we have two outcomes. First, from $p(0,j)$ we obtain:
\begin{equation}
\begin{split}
    p(0,j) &= \frac{1}{4}\big| 1 + P_{ji}\big|^2, \\
            &= \frac{1}{4} (|P_{ji}|^2 + 2 |P_{ji}| + 1),
\end{split}
\end{equation}
From which we set up the following equation
\begin{equation}
    \big| P_{ji}\big|^2 + 2\big| P_{ji}\big| + (1 - 4 p(0,j)) = 0,
\end{equation}
and thus, 
\begin{equation}
\begin{split}
    |P_{ji}| &= \frac{-2 \pm \sqrt{4 - 4(1 - 4p(0,j))}}{2} \\
    &= -1 \pm 2 \sqrt{p(0,j)} = -1 + 2 \sqrt{p(0,j)}
\end{split}
\end{equation}

Similarly, from $p(1,j)$:
\begin{equation}
\begin{split}
    p(1,j) &= \frac{1}{4}\big| 1 - P_{ji}\big|^2, \\
            &= \frac{1}{4} (|P_{ji}|^2 - 2 |P_{ji}| + 1),
\end{split}
\end{equation}
sets up the equation
\begin{equation}
\big| P_{ji}\big|^2 - 2\big| P_{ji}\big| + (1 - 4 p(1,j)) = 0,
\end{equation}
so
\begin{equation}
\begin{split}
    |P_{ji}| &= \frac{2 \pm \sqrt{4 - 4(1 - 4p(1,j))}}{2} \\
    &= 1 \pm 2 \sqrt{p(1,j)} = 1 - 2 \sqrt{p(1,j)}
\end{split}
\end{equation}
In both cases, we have taken the relevant branches of the square root because the expectation value of a unitary operator has to be bounded between $-1$ and 1. Using both, a more consistent result is obtained as 
\begin{equation}
    |P_{ji}| = \sqrt{p(0,j)} - \sqrt{p(1,j)}.
\end{equation}

The change of basis by the unitary $U$, i.e., as in the quantum circuit in Fig.~\ref{fig:modHij_U}, brings an analogous result with a difference in the measurement counts.
\end{itemize}

For evaluating the sign component of the matrix element $|H_{ij}|$, the quantum circuit shown in Fig.~\ref{fig:P_overlap} is utilized. This circuit is of a modified Hadamard test, tailored for determining transition amplitudes rather than expectation values. Its action is as follows: 
\begin{itemize}
\item The initial state is
\begin{equation}
    |\Psi_0\rangle = |0\rangle_a |0\rangle_s.
\end{equation}
\item Apply a Hadamard to the ancilla,
\begin{equation}
    |\Psi_1\rangle
    =
    \frac{1}{\sqrt{2}}
    \left( |0\rangle_a + |1\rangle_a \right)|0\rangle_s.
\end{equation}

\item Apply the controlled-$X$ gates, which generate the computational basis states, using the ancilla qubit as control 
\begin{equation}
    |\Psi_2\rangle
    =
    \frac{1}{\sqrt{2}}
    \left( |0\rangle_a |b_i\rangle_s + |1\rangle_a |b_j\rangle_s \right).
\end{equation}

\item Apply the unitary $U$ on the state register
\begin{equation}
    |\Psi_3\rangle=\frac{1}{\sqrt{2}}\left( |0\rangle_a U |b_i\rangle_s + |1\rangle_a U|b_j\rangle_s \right).
\end{equation}
\item Apply a controlled-$P$ gate on the state register 
\begin{equation}
    |\Psi_4\rangle = \frac{1}{\sqrt{2}} \left( |0\rangle_a P U |b_i\rangle_s + |1\rangle_a U|b_j\rangle_s \right).
\end{equation}
\item Finally, apply a Hadamard gate on the ancilla qubit again
\begin{equation}
\begin{split}
    |\Psi_5\rangle
     =
    &\frac{1}{2}
    ( |0\rangle_a P U |b_i\rangle_s + |1\rangle_a P U |b_i\rangle_s +\\ & + |0\rangle_a U|b_j\rangle_s - |1\rangle_a U|b_j\rangle_s ).
\end{split}
\end{equation}
\item Thus, similarly to the Hadamard test, measuring the ancilla gets bits $0$ and $1$ based on the probabilities
\begin{equation}
\begin{split}
    p(0) &= \frac{1}{2} ( 1 + \Re \langle b_i | U^{\dagger} H U|b_j \rangle ), \\
    p(1) &= \frac{1}{2} ( 1 - \Re \langle b_i | U^{\dagger} H U|b_j \rangle ),
\end{split}
\end{equation}
and the real part of the expectation value is 
\begin{equation}
    \Re \langle b_i | U^{\dagger} H U|b_j \rangle = p(0) - p(1).
\end{equation}
This suffices for the quantum systems studied in this paper, as the Hamiltonians are real, but the imaginary part can be computed as well by application of a phase gate, similar as to the standard Hadamard test. 
\end{itemize}

\end{document}